\DeclareMathAlphabet{\mathscr}{OT1}{pzc}{m}{it}

\documentclass[aps,pra,twocolumn,showpacs,showkeys,amsmath,superscriptaddress,floatfix]{revtex4}

\usepackage{graphicx}
\usepackage{epsfig}
\usepackage{dcolumn}
\usepackage{bm}
\usepackage{amsbsy}
\usepackage{amssymb}
\usepackage[usenames]{color}
\usepackage{mathrsfs}
\usepackage{verbatim}
\usepackage{cancel}
\usepackage{multirow}
\usepackage{tabu}

\begin{document}

\title{Formulation of the twisted-light--matter
interaction at the phase singularity: the twisted-light gauge}

\author{G. F. Quinteiro}

\email{gquinteiro@df.uba.ar}

\affiliation{ Departamento de F\'isica and IFIBA, FCEN, Universidad de Buenos
Aires, Ciudad Universitaria, Pabell\'on I, 1428 Ciudad de Buenos Aires,
Argentina}

\affiliation{Universit\"at M\"unster, Wilhelm-Klemm-Str. 10, 48149 M\"unster,
Germany}

\author{D. E. Reiter}

\affiliation{Universit\"at M\"unster, Wilhelm-Klemm-Str. 10, 48149 M\"unster,
Germany}

\author{T. Kuhn}

\affiliation{Universit\"at M\"unster, Wilhelm-Klemm-Str. 10, 48149 M\"unster,
Germany}

\date{\today}

\begin{abstract}
Twisted light is light carrying orbital angular momentum. The profile of such
a beam is a ring-like structure with a node at the beam axis, where a phase
singularity exists. Due to the strong spatial inhomogeneity the mathematical
description of twisted-light--matter interaction is non-trivial, in
particular close to the phase singularity, where the commonly used
dipole-moment approximation cannot be applied. In this paper we show that, if
the handedness of circular polarization and the orbital angular momentum of
the twisted-light beam have the same sign, a Hamiltonian similar to the
dipole-moment approximation can be derived. However, if the signs differ, in
general the magnetic parts of the light beam become of significant importance
and an interaction Hamiltonian which only accounts for electric fields is
inappropriate. We discuss the consequences of these findings for
twisted-light excitation of a semiconductor nanostructures, e.g., a quantum
dot, placed at the phase singularity.
\end{abstract}



\maketitle

\section{Introduction}

In recent years, there has been intense research work in the topic of highly
inhomogeneous light beams, and in particular, in light carrying orbital
angular momentum (OAM)--also called twisted light (TL) \cite{allen1992orb,
araoka2005int, andrews2011str}. The research in TL spans several areas, such
as the generation of beams \cite{padgett2004lig,woerdemann2009sel}, the
interaction of TL with atoms and molecules
\cite{romero2002qua,babiker2002orb} or with condensed matter
\cite{quinteiro2009the, andersen2006qua, simula2008ang, ueno2009coh,
shigematsu2013orb, watzel2012pho, clayburn2013sea, quinteiro2010ele,
quinteiro2010twi, quinteiro2010bel, quinteiro2009ele1, quinteiro2009ele2,
sbierski2013twi}. TL has already proved to be useful in applications. The
most notable example is perhaps the optical trapping and manipulation of
microscopic particles \cite{padgett2011twe,woerdemann2012adv}. Applications
in other fields are also sought, for example in quantum information
technology, where the OAM adds a new degree of freedom encoding more
information \cite{bozinovic2013ter,
molina2004tri,molina2007twi,dambrosio2013pho}. In addition, theoretical
studies in solid state physics predict, for instance, that TL can induce
electric currents in quantum rings \cite{quinteiro2009ele2}, and new
electronic transitions (forbidden for plane waves) in quantum dots
\cite{quinteiro2009ele1}. This all suggests that TL can be a new powerful
tool to control quantum states in nanotechnological applications.

Two features of TL are particularly striking. First, TL exhibits a vortex or
phase singularity at the beam axis. Second, polarization and OAM are so
intermixed that two beams having the same OAM but opposite circular
polarization behave in a completely different way. This is in contrast to
what happens to plane waves, where the polarization alone does not determine
other important properties. These two features can also be found in other
inhomogeneous beams, namely the so-called azimuthally polarized
\cite{ornigotti2013rad, dorn2003sha, zurita2002mul} fields.

The interaction between TL and matter is particularly interesting due to the
inhomogeneous nature of the TL, and it is worth revisiting its mathematical
formulation. The most general form to describe the light-matter interaction
is the minimal coupling Hamiltonian, where the electromagnetic (EM) fields
enter through their potentials. In many cases of interest, the Hamiltonian
can be rewritten in terms of EM fields using gauge transformations, i.e.,
transformations among potentials that preserve the EM fields
\cite{scully1997qua,cohen1997pho,jackson2002lorenz}. Usually, the
transformations are accompanied by approximations. One of the best-known
among these Hamiltonians is the dipole-moment approximation (DMA). It can be
derived under the assumption that the EM fields vary little in the region
where the matter excitation takes place, and effectively the electric field
$\mathbf E(t)$ is treated as spatially homogeneous. The DMA Hamiltonian then
takes the form $H=- q \mathbf r \cdot \mathbf E(t)=- \mathbf d \cdot \mathbf
E(t)$, where $\mathbf d = q \mathbf r$ is the dipole moment of the material
system.

While gauge invariance is a symmetry property of the electromagnetic
interaction and therefore all observable quantities have to be independent of
the special choice of a gauge, this independence is usually lost when
approximations are performed \cite{cohen1997pho}. A standard example is the
$1s$-$2s$ two-photon transition in a hydrogen atom where it has been
explicitly shown that in the case of the $\mathbf d \cdot \mathbf E$ coupling
already very few intermediate states are sufficient to obtain a very accurate
results while in the case of the $\mathbf p \cdot \mathbf A$ coupling a very
large number of intermediate states is required \cite{bassani1977cho}. A
similar behavior has been found in calculations of the level width of
microwave transitions for the measurement of the Lamb shift
\cite{fried1973vec}. Also for single photon transitions in a H$_2^{+}$ ion
large differences between the two gauges have been found when variational
wave functions for the molecular orbitals are used with, e.g. for the
$2s\sigma$-$2p\pi$ transition, a strong preference of the $\mathbf d \cdot
\mathbf E$ coupling \cite{dalgarno1956app}.

The DMA form of light-matter coupling is advantageous for several additional
reasons: Because the DMA only contains the electric field, it is manifestly
gauge invariant. The momentum operator has a clear physical meaning and can
be used directly for the calculation of quantities like current densities. In
contrast, in the case of the minimal coupling there is an additional,
gauge-dependent contribution, usually called the diamagnetic current
\cite{dressel2002ele}. The difference between the canonical and the
mechanical (or kinetic) momentum may also lead to an apparent ambiguity in
the definition of the photon momentum, as has been discussed in detail in a
recent review by Barnett et al. \cite{Barnett2010the}. Finally, since the DMA
interaction Hamiltonian is linear in the field it can be easily treated
perturbatively while the minimal coupling Hamiltonian contains terms linear
and quadratic in the potential which therefore have to be combined in a
proper way when calculating optical nonlinearities. All these arguments show
that a coupling in terms of the electric field has clear advantages. In fact,
when the light field is sufficiently homogeneous over the size of the matter
system the DMA is perfectly applicable. This holds for example in atomic and
molecular physics and also in the case of nanoscale systems such as quantum
dots, where the matter states are highly localized.

When the inhomogeneous nature of the field becomes important the DMA cannot
be used anymore. One could perform calculations with the minimal coupling
Hamiltonian, which contains the vector potential. Still, it is appealing to
work with a Hamiltonian which contains the electric and magnetic fields only,
because then the theory is manifestly gauge invariant. Of course, using the
so-called Poincar\'e gauge \cite{cohen1997pho, jackson2002lorenz} one can
formally rewrite the Hamiltonian in terms of fields, however, it is not
always possible to express the fields explicitly. A desirable expression
would be one resembling the electric dipole-moment Hamiltonian, but retaining
the spatial dependence: $H=- q \mathbf r  \cdot \mathbf E(\mathbf r,t)$. We
will call this \emph{Electric Field Coupling} (EFC) Hamiltonian. In addition
to the dipole-moment coupling, the EFC Hamiltonian also contains higher order
couplings in the electric field, e.g., quadrupole moments.

In some situations the spatial inhomogeneity of the field can be kept in an
EFC Hamiltonian in a parametric way, while the transition matrix elements are
determined by the coupling via the electric dipole term only
\cite{herbst2003ele, rossi2002the, khitrova1999non, reiter2006con,
reiter2007spatiotemporal}, which has been used to describe for instance
four-wave-mixing phenomena \cite{lindberg1992theory,
leitenstorfer1994excitonic, banyai1995exciton}. Because for TL new
transitions are induced by its OAM \cite{quinteiro2009ele1}, such an approach
would not describe the main feature of TL and, thus, it is crucial to include
the spatial dependence also in the transition matrix elements. Under certain
assumptions, e.g., for the interaction with localized structures placed at
the beam maximum, it is possible to cast the Hamiltonian in a EFC form, and
thus, describe the modified selection rules \cite{koksal2012cha,al2000ato}
for example using a Power-Zienau-Woolley transformation
\cite{babiker2002orb}. However, in this paper we show that for TL-matter
interaction in the vicinity of the beam axis an EFC Hamiltonian cannot in
general be used.

The TL-matter interaction at the phase singularity for highly focused beams
has been analyzed using a multipolar expansion for  electric and magnetic
fields \cite{zurita2002mul, klimov2012mapping}, which already revealed that
higher order electric and magnetic terms can be of significant importance.
Nevertheless, for a subgroup of TL beams we will show that it is possible to
derive an electric multipolar Hamiltonian for the TL-matter interaction close
to the phase singularity, which offers the advantages of a DMA Hamiltonian.

We organize the article as follows. As a next step we briefly revisit the
concepts of gauge transformation  and DMA Hamiltonian necessary to understand
the discussion ahead. In Sec.~\ref{Sec:TL} we introduce the mathematical
representations of TL. In Sec.~\ref{Heur}, using a heuristic derivation much
alike the one found in the literature for the DMA, we arrive at the new
expression for the TL-matter Hamiltonian. Section \ref{Sec:Bfields} shows
that the atypical behavior of the electric and magnetic fields of TL is in
part responsible for the need to modify the Hamiltonian. Section \ref{Lag} is
devoted to a careful derivation of the new Hamiltonian. We wrap up with the
conclusions in Sec.~\ref{conc}. In the appendix we discuss why the EFC gauge,
which seems to be the natural extension of the DMA, is not useful for TL.

\section{Light-matter interaction revisited}
\label{D_LDMA}

The starting point for a mathematical description of the effect of light on
matter is the minimal coupling Hamiltonian, that expresses the external EM
fields in terms of a scalar $U(\mathbf r,t)$ and a vector $\mathbf A(\mathbf
r,t)$ potential. For a single particle of mass $m$ and charge ${q}$ under a
static potential $V(\mathbf r)$, the Hamiltonian reads
\begin{equation}
\label{Eq:Hminimal-coupling}
    H
=
    \frac{1}{2m} [\mathbf p - {q} \, \mathbf A(\mathbf r,t)]^2
    + V(\mathbf r)
    + {q} \, U(\mathbf r,t)
\,.
\end{equation}
It is obtained from the Lagrangian
\begin{eqnarray}
\label{Eq:Lagrange}
   L
&=&
    \frac{1}{2} m \dot{\mathbf r}^2 - V(\mathbf r)
   + {q} \, \dot{\mathbf r} \cdot \mathbf A(\mathbf r,t)
   - {q} \, U(\mathbf r,t)
\,
\end{eqnarray}
via the canonical momentum
\begin{equation}
\label{Eq:canonical}
\mathbf p =\frac{\partial L}{\partial \dot{\mathbf r}} = m \dot{\mathbf r} +
{q} \,  \mathbf A(\mathbf r,t)
\end{equation}
and the Legendre transformation $H={\mathbf p} \cdot \dot{\mathbf r} - L$.

The relationship between potentials and the electric $\mathbf E(\mathbf r,t)$
and magnetic $\mathbf B(\mathbf r,t)$ fields are
\begin{subequations}
\begin{eqnarray}
\label{Eq:EMfields-Pot-E}
   \mathbf E(\mathbf r,t)
&=&
    -\partial_t \mathbf A(\mathbf r,t) - \nabla U(\mathbf r,t)\,,
\\
\label{Eq:EMfields-Pot-B}
   \mathbf B(\mathbf r,t)
&=&
    \nabla \times \mathbf A(\mathbf r,t)
\,.
\end{eqnarray}
\end{subequations}
Gauge transformations are defined such that they preserve the electric and
magnetic fields
\begin{subequations}
\label{Eq:gauge-transf}
\begin{eqnarray}
   \mathbf A'(\mathbf r,t)
&=&
    \mathbf A(\mathbf r,t) + \nabla \chi (\mathbf r,t)\,,\label{Eq:gauge-transf_a}
\\
   U'(\mathbf r,t)
&=&
    U(\mathbf r,t) - \frac{\partial}{\partial t} \chi (\mathbf r,t)
\,,\label{Eq:gauge-transf_u}
\end{eqnarray}
\end{subequations}
where $\chi(\mathbf r,t)$ is the scalar gauge transformation function. Since
the canonical momentum (\ref{Eq:canonical}) depends on the vector potential
it is obviously gauge dependent.

In cases where the EM fields vary little on the scale of the system, taken to
be centered around $\mathbf r=0$, a gauge transformation is sought that would
render $\mathbf A'(\mathbf r,t) = 0$ in the region around $\mathbf r=0$.
Assuming that for external radiation $U(\mathbf r,t)=0$, this is achieved by
the G\"oppert-Mayer gauge transformation $\chi = - \mathbf r \cdot \mathbf
A(0,t)$ \cite{cohen1997pho} leading to the new potentials
\begin{subequations}
\label{Eq:GM-gauge}
\begin{eqnarray}
    \mathbf A' (\mathbf r, t)
&=&
    \mathbf A (\mathbf r, t) -
    \! \mathbf A (0, t)
\nonumber \\
&=& (\mathbf r \cdot \nabla)
    \mathbf A (\mathbf r, t)|_{\mathbf r=0} + \ldots
\\
    U'(\mathbf r, t)
&=&
    - \mathbf r \cdot \mathbf E(0, t)
\,.
\end{eqnarray}
\end{subequations}
By neglecting the derivatives of the old vector potential in the relevant
region of space we can obtain $\mathbf A'(\mathbf r, t) = 0$. This leads to
the well-known DMA Hamiltonian
\begin{eqnarray}\label{Eq:H_DMA}
H = \frac{\mathbf p^2}{2m} + V(\mathbf r) - q \mathbf r \cdot \mathbf E(0,t)
\,,
\end{eqnarray}
which is evidently gauge-invariant. It should be understood that the
requirement $\mathbf A'(\mathbf r,t)=0$ in an extended region of space is a
very stringent one, for it demands the magnetic field to be zero in that
region, in violation to Maxwell's equations for a propagating field.

A striking feature of the DMA is that operators retain their physical
meaning. As an important example, we look at the momentum. Due to the fact
that $\mathbf A'(\mathbf r,t) = 0$, the canonical momentum in the new gauge
is equal to the mechanical momentum $m \dot{\mathbf r}$. The mechanical
momentum is indeed important, since it is a form-invariant operator
\cite{scully1997qua, kobe1978gau}. As such, its eigenvalues are independent
of the gauge and are therefore representing physical quantities. The
canonical momentum, on the other hand, is not form-invariant. This is a
drawback, because the quantized version of the canonical momentum, i.e., the
operator $-i\hbar \nabla$, is typically used to perform calculations
(see sect. IV.A.2b of Ref.~\cite{cohen1997pho}). In order to obtain
measurable quantities such as current densities, the correction due to the
vector potential have to be taken into account, e.g., in terms of a
diamagnetic current \cite{dressel2002ele}. The physical (i.e.,
gauge-independent) current is then given by the sum of two gauge-dependent
contributions.

It is obvious that the DMA cannot be used to describe the interaction of TL
with objects placed close to the beam center because there the electric field
and thus the whole light-matter coupling vanishes. In analogy with the DMA, a
transformation function $\chi = - \mathbf r \cdot \mathbf A(\mathbf r,t)$
could be used \cite{kira1999qua} which keeps the spatial dependence of the
vector potential, yielding new potentials
\begin{subequations}
\label{Eq:local-dipole}
\begin{eqnarray}
   \mathbf A'(\mathbf r,t)
&=&
	- (\mathbf r \cdot \nabla) \mathbf A(\mathbf r,t)
	- \mathbf r \times \mathbf B(\mathbf r,t)
\\
   U'(\mathbf r,t)
&=&
	- \mathbf r \cdot \mathbf E(\mathbf r,t)
\,.
\end{eqnarray}
\end{subequations}
As expected, the new scalar potential has the dipole-like form as that
resulting from the G\"oppert-Mayer transformation, but now with a
position-dependent electric field. Like in the case of the DMA the new vector
potential does not vanish, but it contains spatial derivatives of the old
one. Again, for sufficiently localized charges and a sufficiently smooth
vector potential it may be permissible to disregard these terms resulting in
new potentials $\mathbf A'(\mathbf r,t) \simeq 0$ and $U'(\mathbf r,t)=-
\mathbf r \cdot \mathbf E(\mathbf r,t)$, with the concomitant benefits of
equality of momenta. We call this the Electric Field Coupling (EFC)
approximation. However, we will show below that for TL in the region close
the the beam axis this gauge is not useful.

\section{The vector potential of Twisted light}
\label{Sec:TL}

Let us now come to the case of TL. A TL beam can have different radial
profiles such as Laguerre-Gaussian or Bessel type beams. Here we will
consider the case of a Bessel beam, which has the advantage of being an exact
solution of Maxwell's equations \cite{saleh2007fun}. In mathematical terms,
the vector potential of a monochromatic TL beam in cylindrical coordinates
$\{r, \varphi , z\}$, can be described by $\mathbf A = A_r \mathbf{\hat r} +
A_\varphi \boldsymbol{ \hat \varphi} + A_z \mathbf{\hat z}$ with components
\cite{quinteiro2009the,jauregui2004rot}
\begin{subequations}
\label{Eq:A_Bessel}
\begin{eqnarray}\label{Eq:A_Bessel_r}
    A_r(\mathbf r, t)
&=&
    F_{q_r\ell}(r) \cos[(\omega t-q_z z) -
    (\ell + \sigma)\varphi]\,,
 \\ \label{Eq:A_Bessel_phi}
    A_\varphi(\mathbf r, t)
&=&
    \sigma F_{q_r\ell}(r) \sin[(\omega t-q_z z) - (\ell + \sigma)\varphi]\,,
 \\ \label{Eq:A_Bessel_z}
    A_z(\mathbf r, t)
&=&
    -\sigma \frac{q_r}{q_z} F_{q_r\ell+\sigma}(r)
    \sin[(\omega t-q_z z) - (\ell + \sigma)\varphi]
\,, \nonumber \\
\end{eqnarray}
\end{subequations}
with frequency $\omega$, wave vectors $q_z$ and $q_r$, related by
$q_z^2+q_r^2 = (n\omega/c)^2$, $n$ being the index of refraction of the
medium, and $ \mathbf{\hat r}$, $\boldsymbol{ \hat \varphi}$, $\mathbf{\hat
z}$ denoting unit vectors in cylindrical coordinates.
The integer $\ell$ is related to the OAM of the beam, as will be discussed in
more detail below.
The circular polarization of the field, given by polarization vectors
$\boldsymbol{\epsilon}_{\sigma}= e^{i \sigma \varphi} (\mathbf{ \hat{r}}+ i
\sigma \boldsymbol{ \hat{\varphi}}) = \mathbf{\hat{x}}+i\sigma
\mathbf{\hat{y}}$, is singled out with the variable $\sigma$, which yields
left (right)-handed circular polarization for the values $\sigma = +1(-1)$.
Sometimes, in particular in the quantum theory of light, $\sigma$ is referred
to as the spin angular momentum of the photon with $\sigma\hbar$ being the
spin per photon \cite{VanEnk94}.
The radial profile of the beam $F_{q_r\ell}(r)$ is a Bessel function:
$F_{q_r\ell}(r) = A_0 J_\ell (q_r r)$, with $A_0$ being the amplitude of the
potential. Note that $1/q_r$ is a measure of the beam waist.
The vector potential of Eq.~(\ref{Eq:A_Bessel}) satisfies the Coulomb gauge
condition $\nabla \cdot \mathbf A(\mathbf r, t) = 0$ and the vectorial
Helmholtz equation \cite{volke2002orb}.

\begin{figure}[h]
  \centerline{\includegraphics[scale=.6]{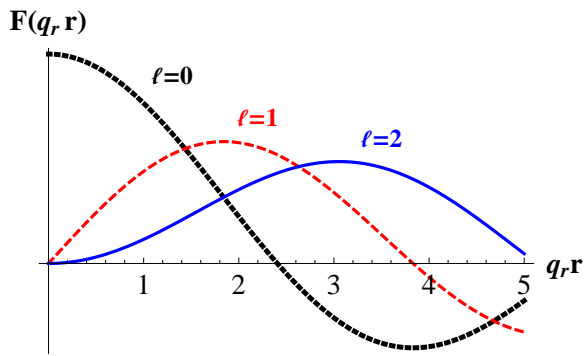}}
  \caption{Beam profiles $F_{q_r\ell}(r)$ of the in-plane components of the
  vector potential (and thus of the in-plane components of the electric field) for a
  non-vortex beam ($\ell = 0$) and TL beams ($\ell =1,2$).}
  \label{fig:Profile}
\end{figure}
Figure \ref{fig:Profile} shows the beam profile $F_{q_r \ell}(r)$ of the
in-plane components $A_r$ and $A_\varphi$ and thus also of the in-plane
components of the electric field for three different values of the OAM:
$\ell= 0, 1, 2$. In the region close to $r = 0$, we observe a main difference
that exists between non-vortex beams and TL. While for non-vortex beams
($\ell = 0$) the amplitude has a maximum at $r = 0$, for TL ($\ell \neq 0$)
the amplitude of the in-plane components is zero there.
Close to the origin, the profile of the in-plane components can be
approximated by $F_{q_r \ell}(r)\propto (q_r r)^{|\ell|}$.

Since for Bessel functions the relation $J_{-\ell} (q_r r) = (-1)^\ell J_\ell
(q_r r)$ holds, it can be seen from Eq.~(\ref{Eq:A_Bessel}) that the
structure of the beam is unchanged if simultaneously the parameters
$(\sigma,\ell)$ are replaced by $(-\sigma,-\ell)$. Therefore in the following
we will restrict ourselves to TL beams with $\ell > 0$.

In the paraxial approximation, when $q_r/q_z\ll 1$, the $z$-component of the
vector potential is disregarded. This case has been extensively used in the
literature \cite{koksal2012cha, babiker2002orb, allen1992orb, andrews2011str,
romero2002qua, watzel2012pho, friese1998opt,nienhuis2004ang,tabosa1999opt}.
The vector potential in the paraxial approximation $\mathbf A^{pa}(\mathbf
r,t)$ then reads
\begin{eqnarray}\label{Eq:Ainitial}
    \mathbf A^{pa}(\mathbf r, t)
&=& A_r(\mathbf r, t) {\hat{\boldsymbol r}}
    + A_\varphi(\mathbf r, t) {\hat{\boldsymbol \varphi}}
\,.
\end{eqnarray}
In this approximation the positive and negative frequency components of
$\mathbf A$ are eigenstates of the angular momentum operator $-i\hbar
\partial /\partial \varphi$  with eigenvalue $\hbar \ell$, which can thus
be identified with the OAM per photon \cite{andrews2011str}. Although this
does not strictly hold in the non-paraxial case, for the sake of brevity, in
the following whenever we refer to the OAM of the beam, we are implicitly
referring to the OAM of its paraxial version. The integer $\ell$ is also
sometimes called the topological charge \cite{andrews2011str}.


\section{A heuristic derivation of the TL-matter interaction}
\label{Heur}

In this section, following the spirit of the DMA, we derive a gauge
transformation that captures the essential features of TL, and at the same
time retains the advantages of the DMA. The derivation is intended to be
intuitive and self-evident, and is only done for the paraxial vector
potential Eq.~(\ref{Eq:Ainitial}). A formal analysis leading to the same
results will be given in Sec.~\ref{Lag}, where the more general form of the
vector potential Eq.~(\ref{Eq:A_Bessel}) will be used and also the
limitations of the paraxial approximation will be discussed.

We are interested in the interaction of TL with a planar, localized
structure, such as a quantum disk or a quantum dot, whose lateral dimensions
are smaller than the the characteristic radial length scale $q_r^{-1}$ of the
beam (i.e., $q_r r\ll 1$). If such a structure is placed at a position with
non-vanishing electric field, in particular close to the beam maximum, the
conditions of the DMA are satisfied and thus the DMA is well applicable.
However, this is different in the case of such a structure centered at $r=0$.
At the beam center the radial profile of the vector potential can be
approximated by $F_{q_r\ell}(r) = \alpha_\ell \, (q_r r)^{\ell}$, with
$\alpha_\ell=A_0 / (2^\ell \ell!)$. Note that the vector potential Eq.\
(\ref{Eq:Ainitial}) and consequently also the electric field at $r=0$ are
zero and thus, within the DMA there would be no interaction whatsoever.

Motivated by the EFC Hamiltonian, we try a gauge transformation function of
the form
\begin{eqnarray}\label{Eq:chi_beta}
    \chi(\mathbf r, t)
&=&
    - \frac{1}{\beta} \mathbf r_\perp \cdot \mathbf A^{pa}(\mathbf r, t)
\,,
\end{eqnarray}
where we have defined a two-dimensional in-plane position vector $\mathbf
r_\perp = r \hat{\mathbf r} = x \hat{\mathbf x} + y \hat{\mathbf y}$ out of
the 3D vector $\mathbf r$, and added a constant prefactor $1/\beta$ to be
determined later. For $\beta = 1$ the transformation obviously reduces to the
EFC case. According to Eqs.~(\ref{Eq:gauge-transf}) the potentials in the new
gauge are calculated to
\begin{subequations}
\begin{eqnarray}
\label{Eq:Heur_Anew}
&&\mathbf A^{pa}\,'(\mathbf r,t)
=
    \left(
        1-\frac{\ell+1}{\beta}
    \right) A^{pa}_r(\mathbf r,t) \,
    \hat{\mathbf r}
\nonumber \\
    &&\mbox{\quad} + \left(
        1-\frac{\sigma\ell+1}{\beta}
    \right) A^{pa}_\varphi(\mathbf r,t) \,
    \hat{\boldsymbol \varphi}
 \\
    &&\mbox{\quad} - \frac{q_z}{q_r}\frac{q_r r}{\beta}  F_{q_r \ell}(r)
        \sin[(\omega t-q_z z) - (\ell + \sigma)\varphi] \,
        \hat{\mathbf z}
\,,
\nonumber \\
    &&U^{pa}\,'(\mathbf r, t)
=
    - \frac{1}{\beta} \mathbf r \cdot \mathbf E^{pa}(\mathbf r, t)
\,.
\end{eqnarray}
\end{subequations}
While the radial dependence of the scalar potential as well as of the
in-plane components of the new vector potential is $\sim (q_r r)^{\ell}$ like
in the case of the old vector potential, the remaining component
$A_z^{pa}\,'$ is $\sim (q_r r)^{\ell+1}$ and can thus be neglected in the
region close to the singularity. This is consistent with keeping terms up to
order $(q_r r)^{\ell}$. In the EFC gauge ($\beta = 1$) we then have
$A_r^{pa}\,'(\mathbf r,t)=-\ell \mathbf A_r^{pa}(\mathbf r,t)$ and
$A_\varphi^{pa}\,'(\mathbf r,t)=-\sigma\ell \mathbf A_\varphi^{pa}(\mathbf
r,t)$. Thus, for $|\ell| \ge 1$ the new vector potential is not smaller than
the old one clearly demonstrating that this gauge does not help to reduce the
difference between mechanical and canonical momentum.

On the other hand, when $\sigma = 1$ the in-plane components
$A_\varphi^{pa}\,'$ and $A_r^{pa}\,'$ of the new vector potential vanish for
$\beta=\ell + 1$. As a result, $\mathbf A^{pa}\,'(\mathbf r,t) = 0$. The
Hamiltonian then reads
\begin{eqnarray}\label{Eq:TL_H_GNA}
H = \frac{\mathbf p^2}{2m}  + V(\mathbf r)
    - \frac{{1}}{\ell + 1} q\mathbf r_\perp \cdot
    \mathbf E^{pa}(\mathbf r, t)
\,.
\end{eqnarray}
We achieve a Hamiltonian which contains an EFC-like term, but with a
different prefactor. Furthermore, since the new vector potential vanishes,
the canonical and mechanical momenta are equal. We will refer to the
transformation according to Eq.~(\ref{Eq:chi_beta}) with $\beta=\ell+1$ as
the {\it TL gauge}.
The very reason for the new prefactor $({\ell+1})^{-1}$ is the existence of a
vortex, that causes the first term of an expansion of the vector potential
near $\mathbf r=0$ to be proportional to $r^{\ell}$.
\begin{figure*}[t]
  \centerline{\includegraphics[scale=.29]{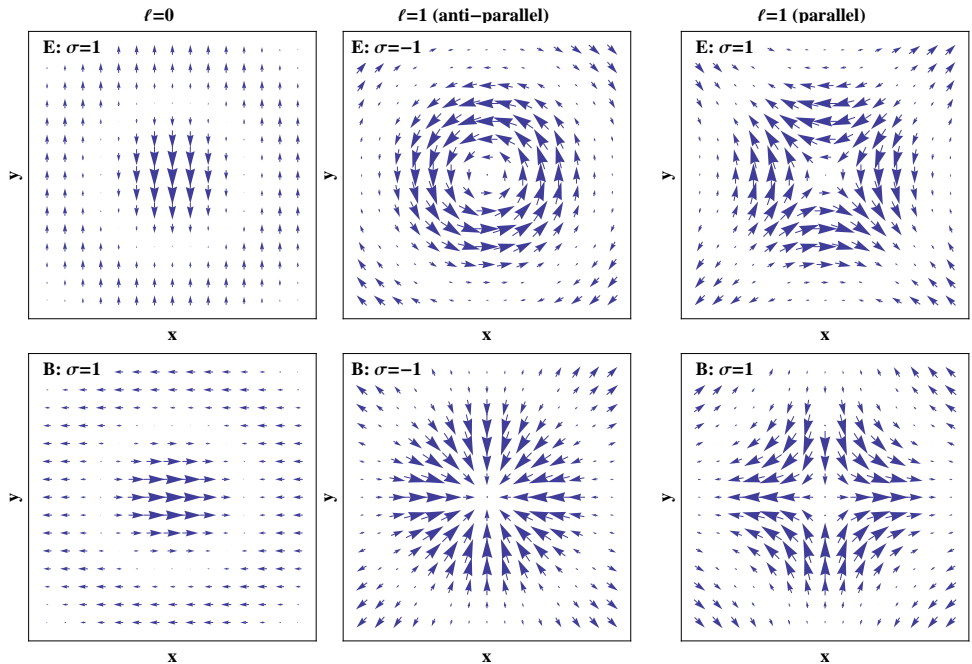}}
  \caption{In-plane components of the electric and magnetic fields at $t=0$ and
  $z=0$ for $\ell=0$ and polarization state $\sigma = +1$ as well as for
  $\ell=1$, and polarization states $\sigma = -1$ (anti-parallel) and
  $\sigma = 1$ (parallel).}
  \label{fig:EFields}
\end{figure*}
For the case $\sigma=-1$, the gauge transformation with $\beta = \ell + 1$ is
not advantageous. This is because $A_\varphi^{pa}\,' \neq 0$, as seen by
inspecting Eq.~(\ref{Eq:Heur_Anew}), and it cannot be neglected. Conversely,
choosing $\beta=-\ell+1$ (for $\ell>1$) would remove the $\varphi$-component
but keep the $r$-component.

It is legitimate to wonder why there is such an asymmetry between TL fields
having the same $\ell$ but differing in their circular polarization state,
while an asymmetry of this type is not present in plane waves. We will
further explore this in the next section.
%

\section{Electric and magnetic fields of TL}
\label{Sec:Bfields}

The aforementioned results suggest that there are two topologically distinct
classes of TL fields, depending on the combination of OAM and circular polarization,
which we will study now in detail.
We calculate the electric and magnetic fields using the full form of the
vector potential [Eq.~(\ref{Eq:A_Bessel})].

A plot of two representative cases of electric and magnetic fields for
$\ell=1$ and $\sigma = \pm 1$, at $t=0$ and $z=0$ is presented in Fig.~\ref{fig:EFields}.
For comparison, the fields of a non-vortex beam with
$\ell=0$ and $\sigma=1$ are also shown. The vectorial character of the
non-vortex beam is similar to a plane wave with perpendicular $\mathbf E$ and
$\mathbf B$ fields. The amplitude is radially modulated according to the
Bessel function $J_0$. In contrast, in the case of TL with $\ell=1$ the field
profiles are much more complex. When $\sigma = -1$, the electric field is
oriented azimuthally around the beam axis, and the magnetic field in the
central region points inwards. For other values of $t$ (or $z$), the patterns
change, but eventually both magnetic and electric fields cycle through
the radially-like and azimuthally-like polarization
patterns. In contrast, when $\sigma = 1$, the fields look entirely different,
and never evolve into azimuthal or radial patterns.
We refer to these two as the anti-parallel [Sign$(\ell)\neq$ Sign$(\sigma)$]
and the parallel [Sign$(\ell)=$ Sign$(\sigma)$] beam classes.

The field patterns shown in Fig.~\ref{fig:EFields} gives an indication why a
gauge, in which the scalar potential provides the dominant contribution to
the coupling could be found in the parallel class but not in the
anti-parallel class. In the central region the field lines of the electric
field in the parallel class are similar to a vector field close to a saddle
point. Such a vector field can indeed be written as the gradient of a scalar
field. In contrast, in the anti-parallel class the field lines of the
electric field are obviously closed indicating that this is dominantly a
vortex-type field which has a non-vanishing curl and therefore cannot be
obtained as a gradient of a scalar potential. Hence, in any gauge the
interaction will mainly originate from the vector potential.

\begin{table*}[t]
\begin{tabular}{|l|l|l|l|}
\hline
\multicolumn{2}{|l|}{\multirow{2}{40pt}{}}
    & \multicolumn{1}{c|}{\bf Parallel}       & \multicolumn{1}{c|}{\bf Anti-parallel} \\
\multicolumn{2}{|l|}{}
    &  \multicolumn{1}{c|}{$\ell>0,\sigma=+1$}   &  \multicolumn{1}{c|}{$\ell>0,\sigma=-1$} \\ \hline
\multirow{3}{45pt}{$\mathbf E/(\alpha_\ell \omega)$}
    & $\hat{\mathbf r}$             & $-(q_r r)^\ell \sin[(\ell+1)\varphi]$                                         & $- (q_r r)^\ell \sin[(\ell-1)\varphi]$ \\ \cline{2-4}
    & $\hat{\boldsymbol \varphi}$   & $- (q_r r)^\ell \cos[(\ell+1)\varphi]$                                        & $ (q_r r)^\ell \cos[(\ell-1)\varphi]$  \\ \cline{2-4}
    & $\hat{\mathbf z}$             & $\frac{1}{2(\ell+1)} \frac{q_r}{q_z} (q_r r)^{\ell+1} \cos[(\ell+1)\varphi]$  & $-2\ell \frac{q_r}{q_z} (q_r r)^{\ell-1}
\cos[(\ell-1)\varphi]$ \\ \hline
\multirow{3}{45pt}{$\mathbf B/(\alpha_\ell q_z)$}
    & $\hat{\mathbf r}$             & \begin{tabular}[c]{@{}l@{}}$ (q_r r)^\ell \left[ 1+\frac{q_r^2}{2q_z^2}  \right] \cos[(\ell+1)\varphi]$\end{tabular}  & \begin{tabular}[c]{@{}l@{}}$- (q_r r)^\ell \left[1+(\ell-1)\frac{q_r^2}{2q_z^2}+ \frac{q_r^2}{2q_z^2} \frac{4\ell (\ell - 1)}{(q_r r)^{2}} \right]$\\ $\times \cos[(\ell-1)\varphi]$\end{tabular} \\ \cline{2-4}
    & $\hat{\boldsymbol \varphi}$   & \begin{tabular}[c]{@{}l@{}}$- (q_r r)^\ell \left[ 1+\frac{q_r^2}{2q_z^2}  \right] \sin[(\ell+1)\varphi]$\end{tabular} & \begin{tabular}[c]{@{}l@{}}$- (q_r r)^\ell \left[1+(\ell+1)\frac{q_r^2}{2q_z^2}-\frac{q_r^2}{2q_z^2} \frac{4\ell (\ell - 1)}{(q_r r)^{2}} \right]$\\ $\times \sin[(\ell-1)\varphi]$\end{tabular} \\ \cline{2-4}
    & $\hat{\mathbf z}$             & $\frac{1}{2(\ell+1)} \frac {q_r}{q_z} (q_r r)^{\ell+1}\sin[(\ell + 1)\varphi]$                                        & \begin{tabular}[c]{@{}l@{}}$ 2\ell \frac{q_r}{q_z}
(q_r r)^{\ell-1} \sin[(\ell-1)\varphi]$ \end{tabular} \\ \hline
\end{tabular}
\caption{\label{Table:Fields} Electric and magnetic field components at $z=0$
and $t=0$ in the region close to the phase singularity calculated from the
full vector potential in Eq.~(\ref{Eq:A_Bessel}).}
\end{table*}
\begin{table*}[t]
\begin{tabular}{|l|l|l|l|}
\hline
\multicolumn{2}{|l|}{\multirow{2}{40pt}{}}
    & \multicolumn{1}{c|}{\bf Parallel}       & \multicolumn{1}{c|}{\bf Anti-parallel} \\
\multicolumn{2}{|l|}{}
    &  \multicolumn{1}{c|}{$\ell>0,\sigma=+1$}   &  \multicolumn{1}{c|}{$\ell>0,\sigma=-1$} \\ \hline
\multirow{3}{45pt}{$\mathbf E^{pa}/(\alpha_\ell \omega)$}
    & $\hat{\mathbf r}$             & $- (q_r r)^\ell \sin[(\ell+1)\varphi]$          & $-(q_r r)^\ell \sin[(\ell-1)\varphi]$ \\ \cline{2-4}
    & $\hat{\boldsymbol \varphi}$   & $- (q_r r)^\ell \cos[(\ell+1)\varphi]$          & $ (q_r r)^\ell \cos[(\ell-1)\varphi]$ \\ \cline{2-4}
    & $\hat{\mathbf z}$             & 0          & 0 \\ \hline
\multirow{3}{45pt}{$\mathbf B^{pa}/(\alpha_\ell q_z)$}
    & $\hat{\mathbf r}$             & \begin{tabular}[c]{@{}l@{}}$ (q_r r)^\ell \cos[(\ell+1)\varphi]$\end{tabular}   & \begin{tabular}[c]{@{}l@{}}$- (q_r r)^\ell \cos[(\ell-1)\varphi]$
\end{tabular} \\ \cline{2-4}
    & $\hat{\boldsymbol \varphi}$   & \begin{tabular}[c]{@{}l@{}}$- (q_r r)^\ell \sin[(\ell+1)\varphi]$\end{tabular}  & \begin{tabular}[c]{@{}l@{}}$- (q_rr)^\ell \sin[(\ell-1)\varphi]$ \end{tabular} \\ \cline{2-4}
    & $\hat{\mathbf z}$             & 0          & 0 \\ \hline
\end{tabular}
\caption{\label{Table:Fields_parax} Same as in Table \ref{Table:Fields}, but
in the paraxial approximation, i.\ e., obtained from
Eq.~(\ref{Eq:Ainitial}).}
\end{table*}

It is our interest to study the region close to the phase singularity $r=0$.
Thus, we provide analytical results for the field amplitudes in this region
expanded in powers of $(q_r r)$. Table \ref{Table:Fields} presents the lowest
non-vanishing orders in $(q_r r)$ of the electric and magnetic fields in the
plane $z=0$ and at $t=0$ obtained from the full vector potential in
Eq.~(\ref{Eq:A_Bessel}). In Table \ref{Table:Fields_parax} the same fields
but obtained from the potential in the paraxial approximation
[Eq.~(\ref{Eq:Ainitial})] are given. Note that the in-plane vector potential
of Eq.~(\ref{Eq:Ainitial}) gives rise to a $z$-component of the magnetic
field; this component, however, has a prefactor $(q_r/q_z)$ and therefore, in
order to be consistent with the paraxial approximation, it has been omitted
in Table \ref{Table:Fields_parax}. Indeed, it is clearly seen that if in
Table \ref{Table:Fields} all terms containing a factor $(q_r/q_z)$ or
$(q_r/q_z)^2$ are neglected the fields of Table \ref{Table:Fields_parax} are
obtained.

Let us first compare the full form and the paraxial case for the parallel
class [Sign($\ell$)$=$Sign($\sigma$), in our case $\sigma=1$]. In the
paraxial approximation we obtain pure in-plane fields with electric and
magnetic fields having the same dependence on $r$. When calculated from the
full vector potential, the magnetic field is slightly rescaled, the
correction being of second order in the small parameter $(q_r/q_z)$. Both
fields acquire a small $z$-component which is of first order in $(q_r/q_z)$.
Additionally it is proportional to $(q_r r)^{\ell+1}$ and thus decreases
faster for $r  \to 0$ than the in-plane components. Thus, these corrections
are negligible in the region close to the phase singularity and the
assumptions of the paraxial approximation are well satisfied in this region.

In the anti-parallel class [Sign($\ell$) $\neq$ Sign($\sigma$), here
$\sigma=-1$], on the other hand, the $z$-components of the electric and
magnetic fields still contain the small parameter $(q_r/q_z)$, however the
radial dependence is now proportional to $(q_r r)^{\ell-1}$. The in-plane
component of the electric field is still proportional to $(q_r r)^{\ell}$,
thus at sufficiently small $r$ the electric field is always dominated by the
$z$-component. This clearly demonstrates that the paraxial approximation,
which neglects the $z$-component, is not applicable in the region close to
the phase singularity. Indeed, a careful look at the $z$-component of the
vector potential in Eq.\ (\ref{Eq:A_Bessel}) reveals that already there the small
factor $(q_r/q_z)$ is counterbalanced by a $r$-dependence which is one order
lower than for the in-plane components and therefore dominates close to
$r=0$.

The dominance of the $z$-component of the fields is closely related to the
field profiles in the anti-parallel case shown in Fig.~\ref{fig:EFields}. As
already mentioned, the electric field profile has a non-vanishing curl which
is oriented in the $z$-direction. According to Maxwell's equation this curl
is associated with the time-derivative of a magnetic field, which therefore
necessarily has to have a strong $z$-component close to $r=0$. Half an
oscillation period later, the roles of electric and magnetic fields in the
central column of Fig.~\ref{fig:EFields} are interchanged. Then the magnetic
field lines are closed circles being associated with a strong $z$-component
of the electric field close to the center.

For angular momenta $\ell \ge 2$  the magnetic field in the anti-parallel
class has an additional correction which is of second order in  $(q_r/q_z)$
but which has a $r$-dependence proportional to $(q_r r)^{\ell-2}$. For
sufficiently small radii this is the dominant contribution to the fields.
Thus, in this case close to the center the beam is dominated by the magnetic
field. This holds in particular for the case $\ell=2$, in which there is a
non-vanishing in-plane magnetic field at the beam center while the electric
field vanishes at this point. This is again an indication that an EFC-like
Hamiltonian is not applicable since with such a Hamiltonian the interaction
with matter is described only in terms of the electric field.

Some research articles in the topics of highly focused TL and
azimuthally/radially-polarized fields report similar findings to ours.
Paraxial beams of TL can be focused using, e.g., high-NA lenses
\cite{iketaki2007inv,bokor2005inv} or a nanoantenna \cite{heeres2014sub}. The
theoretical analysis of focusing, based entirely on electric and magnetic
fields, can be done using the formalism by Wolf \cite{wolf1959ele}, and the
results \cite{iketaki2007inv,bokor2005inv} show important similarities with
the field patterns presented in Fig.~\ref{fig:EFields}.
Azimuthally- and radially-polarized fields are a special class of TL fields
\cite{ornigotti2013rad,volke2002orb}. The field patterns of
azimuthally/radially-polarized {\it non}-paraxial Bessel beams presented by
Ornigotti et al.~\cite{ornigotti2013rad} are also in agreement with our
findings.
Regarding the magnitude of the fields near $r=0$ Zurita-S\'anchez et
al.~\cite{zurita2002mul} have shown that, for the strongly focused
azimuthally-polarized beam they studied, the magnetic interaction overcomes
the electric interaction near the phase singularity; recently, their findings
have been corroborated by the theoretical study of Klimov et
al.~\cite{klimov2012mapping} in the case of focused Laguerre-Gaussian beams.
Finally, in their research on highly-focused TL beams, Monteiro et
al.~\cite{monteiro2009ang}, Iketaki et al.~\cite{iketaki2007inv} and Klimov
et al.~\cite{klimov2012mapping} report that interesting effects only occur
when $\ell=1,2$ and $\sigma=-1$.
The overall similarities are no coincidence, for the vector potential
Eq.~(\ref{Eq:A_Bessel}) --in contrast to Eq.~(\ref{Eq:Ainitial})-- shares
with the aforementioned {\it non}-paraxial beams the important feature of
having a non-negligible $z$-component,  which we have shown to give rise to
the described features.

We are now in a position to clarify the findings in the heuristic derivation
of the TL-matter coupling shown in Sec.~\ref{Heur}. There it was assumed that
there is no $z$-component in the vector potential. From Table
\ref{Table:Fields} we see that the $z$-component of the fields are negligible
only in the parallel class. In the anti-parallel class they are proportional
to $(q_r r)^{-1} \,F_{q_r \ell}(r)$. Because for $r \rightarrow 0$ the
magnetic field cannot be neglected compared to the electric field, we were
not able to derive an EFC-like Hamiltonian.
In other words a Hamiltonian representation given solely in terms of the
electric multipoles, such as $-(1/\beta) q\mathbf r \cdot \mathbf E (\mathbf
r, t)$, is insufficient to describe the TL-matter interaction for the anti-parallel class.

\section{Formal derivation of the TL-matter interaction}
\label{Lag}

The use of the gauge transformation function $\chi(\mathbf r, t)$ found in
Sec.~\ref{Heur} can be motivated using formal arguments. In the following we
use the more general form of Eq.~(\ref{Eq:A_Bessel}) for the vector potential
in the Coulomb gauge.

For charged particles localized around the same center, a
Power-Zienau-Woolley (PZW) transformation can be done using the gauge
function
\begin{eqnarray}\label{Eq:PZW}
    \chi(\mathbf r, t) = - \int_0^1 \mathbf r \cdot \mathbf A(u \mathbf r, t) du
\,,
\end{eqnarray}
where $\mathbf A(u \mathbf r, t)$ is given in the Coulomb gauge. This is the
generalization to inhomogeneous fields of the G\"oppert-Mayer transformation
(DMA), and leads to the so-called Poincar\'e gauge
\cite{cohen1997pho,jackson2002lorenz}.
In our work we focus on the interaction of TL with planar systems. Therefore,
if we consider a charge distribution mainly extended in the $x-y$ plane for a
fixed $z$, the quantity $u \mathbf r$ scales only in the in-plane component
with $u \mathbf r \simeq (u r,\varphi, z)$ (see, e.g.,
Ref.~\cite{cohen1997pho}).

Defining $\mathbf r = \mathbf r_\perp + z \mathbf{\hat z} =r \hat{\mathbf r}
+ z \mathbf{\hat z}$, the gauge function reads
\begin{eqnarray}\label{Eq:chi_Poincare}
   \chi(\mathbf r, t)
&=&
   - \int_0^1 \mathbf r_\perp \cdot
        \mathbf A(u r, \varphi,z; t) du
\nonumber \\
&&
   - \int_0^1 z A_z(u r, \varphi, z; t) du
\,.
\end{eqnarray}
For small systems ($q_r r \ll 1$) the radial dependence can be approximated
by $F_{q_r \ell}(r)\simeq \alpha_\ell (q_r r)^{|\ell|}$, which leads to
$F_{q_r \ell}(u r) \simeq u^{|\ell|} F_{q_r \ell}(r)$. With these
simplifications, we evaluate the integrals Eq.~(\ref{Eq:chi_Poincare}), and
obtain
\begin{eqnarray}\label{Eq:chiPZW}
    \hspace{-4mm}
    \chi(\mathbf r, t)
&=&
    - \frac{1}{|\ell|+1}  \mathbf r_\perp \! \cdot \!
    \mathbf A(\mathbf r,t)
\nonumber \\
&&
    - \frac{1}{|\ell+\sigma|+1} z A_z(\mathbf r,t)
\,.
\end{eqnarray}
The in-plane part of the transformation function $\chi(\mathbf r, t)$ is
exactly the same as we got in Sec.~\ref{Heur}. In addition, there is a new
term arising from the non-vanishing $z$-component of the vector potential.
Note that we have neither required $\mathbf A' (\mathbf r, t) = 0$ in the new
gauge, nor have we neglected $A_z(\mathbf r, t)$. Additionally, for
non-vortex fields ($\ell = 0$) with negligible $z$-component our result
coincides with that of the EFC Hamiltonian.

Here we have motivated the use of Eq.~(\ref{Eq:chiPZW}) by showing that the
TL gauge function can formally be derived by a PZW transformation for charged
particles localized in a planar structure (constant $z$). Since  any gauge transformation function can be postulated and used to cast the potential in suitable forms, the TL gauge can also be applied
to other  more general structures for variable $z$  (with
varying degrees of accuracy or usefulness).

We see that the natural extension of the DMA to the case of TL beams is
slightly different from the plain EFC Hamiltonian. Because of the generalized
use of EFC Hamiltonians \cite{koksal2012cha,al2000ato,babiker2002orb}, it is
worth exploring further its connection to our result. To this end, let us
simply postulate a general gauge transformation of the form
\begin{eqnarray}\label{Eq:chiGeneral}
    \chi_\beta (\mathbf r, t)
&=&
    - \frac{1}{\beta_r}  \mathbf r_\perp \cdot
    \mathbf A(\mathbf r,t)
    - \frac{1}{\beta_z}  z  A_z(\mathbf r,t)
\,,
\end{eqnarray}
where $\beta_i$ is any number. Clearly, we can recover the TL gauge by
$\beta_r=|\ell| +1$ and $\beta_z=|\ell+\sigma| +1$. In contrast, when setting
$\beta_i=1$, Eq.~(\ref{Eq:chiGeneral}) reduces to the EFC gauge. In the
following, we will again only consider the case of $\ell>0$ and polarization
$\sigma = \pm 1$, since, as already discussed, there are no essential
differences in the case with negative $\ell$ and opposite sign of $\sigma$.

According to Eqs.~(\ref{Eq:gauge-transf_u}) and (\ref{Eq:chiGeneral}), the
scalar potential in the new gauge reads
\begin{eqnarray}\label{Eq:UPGeneral}
    U'(\mathbf r, t)
&=&
    - \frac{1}{\beta_r}  \mathbf r_\perp \cdot \mathbf
    E(\mathbf r,t)
    - \frac{1}{\beta_z}  z  E_z(\mathbf r,t)
\,.
\end{eqnarray}
Obviously in the scalar potential we recover an EFC-type structure of the
Hamiltonian, however with in general different prefactors for the in-plane
and out-of-plane components. The new vector potential in the region close to
the phase singularity is given by
\begin{widetext}
\begin{subequations}
\label{Eq:New_A}
\begin{eqnarray}
 {{A'}_r} &=& \frac{{{\beta _r} - \left( {1 + \ell } \right)}}{{{\beta _r}}}{A_r}
 + \frac{\sigma}{{2{\beta _z}}} \frac{{q_r^2}}{{q_z^2}}{A_0}
 {\frac{1}{{\left( {\ell  + \sigma } \right)!}}}
 {{\left( {\frac{{{q_r}r}}{2}} \right)}^{\ell  + \sigma  - 1}} \left({q_z}z\right) \nonumber \\
 &&\mbox{\qquad}\times  \left[{\left( {\ell  + \sigma } \right)}
 - \frac{{\left( {\ell  + \sigma  + 2} \right)}}{{\left( {\ell  + \sigma  + 1} \right)}}
 {{\left( {\frac{{{q_r}r}}{2}} \right)}^{2}} \right]
 \sin \left[ {\left( {\omega t - {q_z}z} \right) - \left( {\ell  + \sigma } \right)\varphi } \right]\,, \label{Eq:New_A_r} \\
{{A'}_\varphi } &=& \frac{{{\beta _r} - \left( {1 + \sigma \ell } \right)}}{{{\beta _r}}}
 {A_\varphi } - \frac{\sigma}{{{2\beta _z}}} \frac{{q_r^2}}{{q_z^2}}{A_0}
 \frac{{\left( {\ell  + \sigma } \right)}}{{\left( {\ell  + \sigma } \right)!}}
 {{\left( {\frac{{{q_r}r}}{2}} \right)}^{\ell  + \sigma  - 1}} \left({q_z}z\right) \nonumber \\
 &&\mbox{\qquad}\times  \left[ 1 - \frac{1}{{\left( {\ell  + \sigma  + 1} \right)}}
 {{\left( {\frac{{{q_r}r}}{2}} \right)}^{2}} \right]
 \cos \left[ {\left( {\omega t - {q_z}z} \right) - \left( {\ell  + \sigma } \right)\varphi } \right]\,, \label{Eq:New_A_vp}\\
{{A'}_z} &=& \frac{{{\beta _z} - 1}}{{{\beta _z}}}{A_z}
 - \frac{\sigma}{{{\beta _z}}} \frac{{{q_r}}}{{{q_z}}}{A_0}
 \frac{1}{{\left( {\ell  + \sigma } \right)!}}
 {{\left( {\frac{{{q_r}r}}{2}} \right)}^{\ell  + \sigma }} \left({q_z}z\right) \nonumber \\
 &&\mbox{\qquad}\times \left[ 1 - \frac{1}{{\left( {\ell  + \sigma  + 1} \right)}}{{\left( {\frac{{{q_r}r}}{2}} \right)}^{2}} \right]
 \cos \left[ {\left( {\omega t - {q_z}z} \right) - \left( {\ell  + \sigma } \right)\varphi } \right] \nonumber \\
 &&- \frac{{2}}{{{\beta _r}}}\frac{{{q_z}}}{{{q_r}}}{A_0}\frac{1}{{\ell !}}
 {\left( {\frac{{{q_r}r}}{2}} \right)^{\ell  + 1}}
 \sin \left[ {\left( {\omega t - {q_z}z} \right) - \left( {\ell  + \sigma } \right)\varphi } \right]\,,\label{Eq:New_A_z}
\end{eqnarray}
\end{subequations}
\end{widetext}
where we have used the expansion in powers of $(q_r r)$ and kept all terms up
to the order $\ell+1$.

As discussed in Sec.~\ref{Heur} for the paraxial approximation, also here it
is obvious that the EFC gauge with $\beta_i = 1$ is not useful for TL because
the transformed in-plane components of the vector potential are not smaller
than the original ones. In fact, for $\ell > 1$ they are in general even
larger. A more detailed discussion of the EFC gauge can be found in the
appendix. In the following we will restrict ourselves to the TL gauge
$\beta_r=\ell +1$ and $\beta_z=\ell+\sigma +1$ and discuss the vector
potential for the different cases. We remind that $A_{r}(\mathbf r,t)$ and
$A_{\varphi}(\mathbf r,t)$ are proportional to $(q_r r)^{\ell}$ while
$A_{z}(\mathbf r,t) \propto (q_r r)^{\ell+\sigma}$.

\subsection{Vector potential in the parallel class}
\label{Sec:parallel}

We first examine the new vector potential in the parallel class, i.e.,
Sign($\ell$) $=$ Sign($\sigma$) or, more explicitly, $\sigma = 1$. The
results are a direct extension to those found by the heuristic derivation in
Sec.~\ref{Heur}. In the parallel class the radial dependence of the
transformed potential is the same as for the original one. Moreover, each
component of the vector potential contains a term proportional to the small
quantities $(q_z z)$. Since we assume a planar structure these terms can be
neglected. Then the expressions simplify to
\begin{subequations}
\begin{eqnarray}
\label{Eq:APbeta2_r}
    A_r'(\mathbf r,t)
&=&
    0\,,
\\
\label{Eq:APbeta2_varphi}
    A_\varphi'(\mathbf r,t)
&=&
    0\,,
\\
\label{Eq:APbeta2_z}
    A_{z}'(\mathbf r,t)
&=&
    \left[
        \frac{1+\ell}{2+\ell}
        + 2 \left(\frac{q_z}{q_r}\right)^2
    \right]
    A_{z}(\mathbf r,t)
\,.
\end{eqnarray}
\end{subequations}
The first thing to notice is that the components $A_r'$ and $A_\varphi'$ of
the new vector potential are zero, as we have already found in
Sec.~\ref{Heur}. Therefore, in the Hamiltonian
\begin{eqnarray}\label{Eq:H_Parallel}
    H
&=&
    \frac{{\mathbf p}^2}{2m}  + V(\mathbf r)
    - \frac{{1}}{\ell + 1} q \mathbf r_\perp \cdot
    \mathbf E(\mathbf r,t)
    - \frac{{1}}{\ell + 2} q z E_z(\mathbf r,t)
\nonumber \\
&&
    - \frac{{q}}{m}
    p_z A'_z(\mathbf r,t)
    + \frac{{q}^2}{2m}
    A'_z\,^2(\mathbf r,t)
\,,
\end{eqnarray}
the in-plane TL-matter interaction can be expressed solely by a dipole-like
term $ - (\ell + 1)^{-1} q \mathbf r_\perp \cdot    \mathbf E(\mathbf r,t)$
with a prefactor different to the EFC Hamiltonian. For the $z$-component we
have both a dipole-like term, but also a term $- (q/m) p_z A'_z(\mathbf r,t)$
which still contains the vector potential.
We point out that $[{q}^2/(2m)] A'_z\,^2(\mathbf r,t) \propto (q_r
r)^{2\ell+2}$, and may be safely disregarded.

It is interesting to also compare again the canonical and mechanical momenta.
Their difference is given by
\begin{eqnarray}\label{Eq:CanonicalP_v2}
    \mathbf p - m \dot{\mathbf r}
=
    q A_{z}'(\mathbf r,t) \, {\hat{\mathbf z}}
\,.
\end{eqnarray}
Also here the canonical and mechanical momenta are the same for the in-plane
components and only in the $z$-component a difference in the momenta arises,
which is however of the order $(q_r r)^{\ell+1}$ and therefore one order
higher than the correction to the in-plane momenta in the original gauge.

Let us now consider what happens in situations of experimental and
application interest.
We first address the situation when the interaction with the system only
occurs through the in-plane components of the field, for example in the
selective excitation of heavy holes in a quantum dot. Then, the TL-matter
interaction reads $H_{\mathrm{TL-matter}} =  - (\ell + 1)^{-1}
q\mathbf{r}_\perp \cdot \mathbf E(\mathbf r,t)$ and is modeled by electric
multipoles only with all the benefits of a DMA. Effectively we end up in the
desirable situation where the vector potential is eliminated, as also shown
in Sec.~\ref{Heur}. Nevertheless the description is beyond the DMA because it
keeps the full spatial dependence of the electric field and thus can give
rise to transitions which are forbidden in the case of excitation by plane
waves, for example transitions from envelope function with $s$-type symmetry
in the valence band to those with $p$-type symmetry in the conduction band or
vice versa.

Next, we consider the case where the system interacts with the $z$-component
of the field, for example in intersubband transitions in quantum wells
\cite{sbierski2013twi} or the excitation of light holes
\cite{quinteiro2014lig}. Here, the electric multipoles are accompanied by a
magnetic term arising from the non-vanishing $z$-component of the vector
potential.
However, since no atypical behavior of the fields near the phase singularity
occurs, it is expected that the electric interaction is larger than the
magnetic one as usually happens. One could then safely only retain the
electric multipolar term, and possibly neglect the difference between
momenta.
Therefore for the parallel class a Hamiltonian with only electric
dipole-moment terms having the correct prefactors can describe the TL-matter
interaction at the phase singularity.

\subsection{Vector potential in the anti-parallel class}
\label{Sec:Antiparallel}

For the anti-parallel class, we already found that a description with
electric field only is not sufficient. Still, we can gain valuable insights
from studying the anti-parallel case with Sign($\ell$) $\neq$ Sign($\sigma$),
i.e., $\sigma = -1$. Here, the vector potential reads:
\begin{widetext}
\begin{subequations}
\begin{eqnarray} \label{Eq:APbeta4_r}
    A_r'(\mathbf r,t)
&=&
    - \frac{q_r^2}{2q_z^2} \frac{q_z z}{\ell} \left[ (\ell+1) -
\frac{4\ell(\ell-1)}{(q_r r)^2}\right] A_\varphi(\mathbf r,t) \,,
\\
\label{Eq:APbeta4_varphi}
    A_\varphi'(\mathbf r,t)
&=&
    \frac{2\ell}{\ell+1} A_\varphi(\mathbf r,t)
    - \frac{q_r^2}{2q_z^2} \frac{q_z z}{\ell} \left[ (\ell-1) -
\frac{4\ell(\ell-1)}{(q_r r)^2}\right] A_r(\mathbf r,t) \,,
\\
\label{Eq:APbeta4_z}
    A_{z}'(\mathbf r,t)
&=&
    \left[ \frac{\ell-1}{\ell} - \frac{2q_z}{q_r}\right] A_z(\mathbf r,t)
    - \frac{q_r}{q_z} \frac{q_z z}{\ell} \left[ \frac{q_r r}{2} -
\frac{2\ell}{q_r r}\right] A_r(\mathbf r,t)
\,.
\end{eqnarray}
\end{subequations}
\end{widetext}
Here we have kept the terms $\propto (q_z z)$ since, in contrast to the
parallel class, they are now accompanied by radial dependencies proportional
to $(q_r r)^{-1}$ and $(q_r r)^{-2}$ times the original vector potential.
Thus, the transformed vector potential becomes even stronger close to the
phase singularity. The magnetic interaction resulting from these terms may be
comparable or even surpass the electric interaction. This is in agreement
with previous results for highly focused beams, where a magnetic field
contribution stronger than the electric field contribution at the phase
singularity was found \cite{zurita2002mul, klimov2012mapping}. It is also
interesting, that even far from the phase singularity, the in-plane term
$A_\varphi'$ does not vanish.

Let us study this in more detail using as an example the excitation of a
quantum dot placed at the beam axis by a TL beam and energy close to the QD
band-gap. Considering again the case of optical transitions with in-plane
matrix elements such as the heavy hole-to-conduction band transitions, we
neglect the $z$-component of the interaction, and also the terms proportional
to $A'(\mathbf r,t)^2$. Then, the Hamiltonian reduces to
\begin{eqnarray}\label{Eq:H_anti-Parallel}
    H
&\simeq&
    \frac{{\mathbf p}^2}{2m}  + V(\mathbf r)
        - \frac{{1}}{\ell + 1} q \mathbf r_\perp \cdot
    \mathbf E(\mathbf r,t)
\nonumber \\
&&
    - \frac{{q}}{2 m} \left[
     \mathbf p_\perp \cdot \mathbf A'_\perp(\mathbf r,t)
    +  \mathbf A'_\perp(\mathbf r,t) \cdot \mathbf p_\perp
    \right]
\,.
\end{eqnarray}
[We note that the angular component of the momentum vector reads $(\mathbf
p)_\varphi = (1/r) p_\varphi$, where the canonical momentum $p_\varphi = \partial L/\partial \dot{\varphi}$ is in fact an angular momentum \cite{goldstein1962cla}.]
Though there is an EFC-type Hamiltonian, clearly the in-plane vector
potential remains in the Hamiltonian. We wonder how electric and magnetic
contributions compare to each other. Let us specifically consider the case
$\ell=2$. Then, the electric multipolar term is proportional to $r (q_r
r)^2$. On the other hand, the magnetic term in Eq.\ (\ref{Eq:H_anti-Parallel}) is
proportional to $p (q_r r)^{0}$. If we assume that momentum and position
vector are proportional to each other, as it is so in the DMA (since $\mathbf
p=-i (m/\hbar)[\mathbf r,H_0]$), it becomes clear that one should not a
priori neglect the magnetic interaction, for it may be comparable or even
larger the electric interaction, in particular at the phase singularity.

When the $z$-component of the fields become also important, it is clear that
also here the vector potential remains in the Hamiltonian. Thus, for the
anti-parallel class the TL-gauge transformation, though being mathematically
correct, is in general not advantageous.

%
%


\section{Conclusions}
\label{conc}

We have studied the TL-matter interaction close to the beam axis. In contrast
to conventional light beams, twisted light has a phase singularity at the
point $r=0$, and a strong intermixing between polarization and OAM. We
distinguished the TL beams into two topologically different classes, namely
the parallel class where handedness of circular polarization (i.e., the
photon spin) and OAM have the same sign and the anti-parallel class where the
signs of circular polarization and OAM differ.

To obtain a Hamiltonian which includes the EM fields instead of the
potentials, we suggested to use a new gauge, the TL gauge. For the parallel
class, the TL gauge leads to a Hamiltonian which has a dipole-type structure,
but a different prefactor. For in-plane problems it takes the simple form
$H_{\mathrm{TL-matter}} =  - (|\ell | + 1)^{-1} q \mathbf{r}_\perp \cdot
\mathbf E(\mathbf r,t)$. The prefactor is mandatory to describe the correct
interaction and to achieve the identity of canonical and mechanical momentum.
The origin of the prefactor in the TL gauge is the vortex, which exists at
the phase singularity. For the anti-parallel class we showed that the TL
gauge, which casts the Hamiltonian at least partly into electric fields, is
not in general advantageous as the vector potential cannot be eliminated nor
neglected. Because in the anti-parallel class magnetic effects cannot be
neglected compared to the electric ones, the Hamiltonian should include
magnetic as well as electric terms, and their relative strength must be
analyzed in the particular problem at hand.

We compared the TL gauge to the more common DMA and EFC Hamiltonians. While
for structures located close to the beam maximum the DMA is applicable, for
structures located close to the beam center it cannot be used since the
electric field at the phase singularity vanishes.
We have also pointed out that the use of the paraxial approximation close to
the phase singularity may be misleading and should be avoided at least in the
anti-parallel beam class.

In contrast to other gauges the TL gauge depends explicitly on beam
parameters, in particular on the OAM $\ell$. On the one hand this is clearly
a restriction, but on the other hand, when TL is used to excite structures in
the region of the beam center, this is usually done with the aim to address
specific transitions which are driven by a light field with a given value of
$\ell$. In this case a beam with a well-defined $\ell$ is used and thus, at
least for beams within the parallel class, the TL gauge for this experimental
set-up is well defined and can be used to write the coupling completely in
terms of the electric field.

In comparison to other gauges, like the Poincar\'e-gauge or the multipolar
gauge, the TL gauge offers the same advantages for TL beams in the parallel
class as the DMA offers for slowly varying light beams: In contrast to the
Poincar\'e-gauge, the TL gauge can be simply evaluated and leads to explicit
formulas. For in-plane problems $H_{\mathrm{TL-matter}}$ contains only the
electric field, which makes it manifestly gauge invariant and secures the
physical meaning of the momentum operator. Furthermore, it contains all the
higher order electric field couplings like coupling to quadrupole terms in a
compact, appealing form.

\section{Acknowledgment}

G.~F.~Quinteiro would like to thank P.~I.~Tamborenea for fruitful discussions
in the topic of gauge-invariance in general and applied to TL. He also thanks
the Argentine research agency CONICET for financial support through the
``Programa de Becas Externas''.

\appendix

\section{EFC gauge for twisted light} \label{Sec:parallel_LDMA}

In this appendix we will discuss in some more detail why the EFC gauge is not
useful for TL. We obtain the EFC gauge from our more general gauge function
Eq.~(\ref{Eq:chiGeneral}) by setting $\beta_r=\beta_z = 1$. From the general
formulas (\ref{Eq:New_A}) we can then calculate the new potentials. In the
parallel class, when restricting ourselves to the lowest non-vanishing order
in $(q_r r)$ (which is the order $\ell$ for the in-plane components and
$\ell+1$ for the $z$-component) and, as discussed in Sec.~\ref{Sec:parallel},
neglecting terms involving the small quantities $\propto (q_z z)$, the
expressions simplify to
\begin{subequations}
\label{Eq:APbeta1}
\begin{eqnarray} \label{Eq:APbeta1_r}
    A_r'(\mathbf r,t)
&=&
    -\ell
   A_{r}(\mathbf r,t)\,,
\\
\label{Eq:APbeta1_varphi}
    A_\varphi'(\mathbf r,t)
&=&
    -\ell
   A_{\varphi}(\mathbf r,t)\,,
\\
\label{Eq:APbeta1_z}
    A_{z}'(\mathbf r,t)
&=&
    2(1+\ell) \left(\frac{q_z}{q_r}\right)^2
    A_{z}(\mathbf r,t)
\,.
\end{eqnarray}
\end{subequations}
We see that the vector potential in the EFC gauge grows with $\ell$.  At
first glance the Eqs.~(\ref{Eq:APbeta1}) might look surprising since they
seem to violate the uniqueness of the EM fields: If one wanted to calculate
the z-component of the magnetic  field using the in-plane components of the
new vector potential, one would find that $B_{z}'=(\nabla \times \mathbf
A')_z=-\ell (\nabla \times \mathbf A)_z = -\ell B_z$ which would violate the
independence of the EM field on gauge transformations. Such a contradiction
is only apparent for the following reason. We approximated the in-plane
components of the vector potential to lowest order in $q_r r$, i.e., $(q_r
r)^{\ell}$. Under this approximation $B_z=(\nabla \times \mathbf A)_z=0$ and
thus there is no contradiction.

The transformed vector potential also reveals a difference between canonical and mechanical
momentum  $\mathbf p - m \dot{\mathbf r} = {q} \mathbf A'(\mathbf r,t)$,
which also grows with $\ell$. Therefore, when the EFC gauge is applied to
high-$\ell$ TL beams at the phase singularity and the canonical momentum
instead of the mechanical momentum is used in calculations, a significant
error may be introduced.

In the anti-parallel class we have
\begin{subequations}
\begin{eqnarray}
\label{Eq:APbeta3_r}
&&	A_r'(\mathbf r,t)
=
	-\ell A_r(\mathbf r,t) \nonumber \\
&&\mbox{\quad}	- \frac{q_r^2}{2q_z^2} (q_z z) \left[ (\ell+1)
- \frac{4\ell(\ell-1)}{(q_r r)^2}\right]
A_\varphi(\mathbf r,t)\,, \mbox{\qquad}
\\
\label{Eq:APbeta3_varphi}
&&	A_\varphi'(\mathbf r,t)
=
	\ell A_\varphi(\mathbf r,t) \nonumber \\
&&\mbox{\quad}	- \frac{q_r^2}{2q_z^2} (q_z z) \left[ (\ell-1)
- \frac{4\ell(\ell-1)}{(q_r r)^2}\right]
A_r(\mathbf r,t)\,,
\\
\label{Eq:APbeta3_z}
&&	A_{z}'(\mathbf r,t)
=
	- \frac{2q_z}{q_r} (\ell+1) A_z(\mathbf r,t) \nonumber \\
&&\mbox{\quad}	- \frac{q_r}{q_z} (q_z z)\left[ \frac{q_r r}{2} -\frac{2\ell}{q_r r}\right]
A_r(\mathbf r,t)
\,.
\end{eqnarray}
\end{subequations}
Again, the in-plane components grow with increasing $\ell$. Furthermore, like
in the case of the TL gauge in the anti-parallel class
(Sec.~\ref{Sec:Antiparallel}), the vector potential exhibits new terms
containing $(q_r r)^{-n}$ multiplying the original vector potential. Thus,
also in the EFC gauge the transformed vector potential becomes even stronger
close to the phase singularity. For both reasons the EFC gauge is not useful
in the anti-parallel class.
%



\begin{thebibliography}{61}%
\makeatletter
\providecommand \@ifxundefined [1]{%
 \@ifx{#1\undefined}
}%
\providecommand \@ifnum [1]{%
 \ifnum #1\expandafter \@firstoftwo
 \else \expandafter \@secondoftwo
 \fi
}%
\providecommand \@ifx [1]{%
 \ifx #1\expandafter \@firstoftwo
 \else \expandafter \@secondoftwo
 \fi
}%
\providecommand \natexlab [1]{#1}%
\providecommand \enquote  [1]{``#1''}%
\providecommand \bibnamefont  [1]{#1}%
\providecommand \bibfnamefont [1]{#1}%
\providecommand \citenamefont [1]{#1}%
\providecommand \href@noop [0]{\@secondoftwo}%
\providecommand \href [0]{\begingroup \@sanitize@url \@href}%
\providecommand \@href[1]{\@@startlink{#1}\@@href}%
\providecommand \@@href[1]{\endgroup#1\@@endlink}%
\providecommand \@sanitize@url [0]{\catcode `\\12\catcode `\$12\catcode
  `\&12\catcode `\#12\catcode `\^12\catcode `\_12\catcode `\%12\relax}%
\providecommand \@@startlink[1]{}%
\providecommand \@@endlink[0]{}%
\providecommand \url  [0]{\begingroup\@sanitize@url \@url }%
\providecommand \@url [1]{\endgroup\@href {#1}{\urlprefix }}%
\providecommand \urlprefix  [0]{URL }%
\providecommand \Eprint [0]{\href }%
\providecommand \doibase [0]{http://dx.doi.org/}%
\providecommand \selectlanguage [0]{\@gobble}%
\providecommand \bibinfo  [0]{\@secondoftwo}%
\providecommand \bibfield  [0]{\@secondoftwo}%
\providecommand \translation [1]{[#1]}%
\providecommand \BibitemOpen [0]{}%
\providecommand \bibitemStop [0]{}%
\providecommand \bibitemNoStop [0]{.\EOS\space}%
\providecommand \EOS [0]{\spacefactor3000\relax}%
\providecommand \BibitemShut  [1]{\csname bibitem#1\endcsname}%
\let\auto@bib@innerbib\@empty
\bibitem [{\citenamefont {Allen}\ \emph {et~al.}(1992)\citenamefont {Allen},
  \citenamefont {Beijersbergen}, \citenamefont {Spreeuw},\ and\ \citenamefont
  {Woerdman}}]{allen1992orb}%
  \BibitemOpen
  \bibfield  {author} {\bibinfo {author} {\bibfnamefont {L.}~\bibnamefont
  {Allen}}, \bibinfo {author} {\bibfnamefont {M.~W.}\ \bibnamefont
  {Beijersbergen}}, \bibinfo {author} {\bibfnamefont {R.~J.~C.}\ \bibnamefont
  {Spreeuw}}, \ and\ \bibinfo {author} {\bibfnamefont {J.~P.}\ \bibnamefont
  {Woerdman}},\ }\href@noop {} {\bibfield  {journal} {\bibinfo  {journal}
  {Phys.\ Rev.\ A}\ }\textbf {\bibinfo {volume} {45}},\ \bibinfo {pages} {8185}
  (\bibinfo {year} {1992})}\BibitemShut {NoStop}%
\bibitem [{\citenamefont {Araoka}\ \emph {et~al.}(2005)\citenamefont {Araoka},
  \citenamefont {Verbiest}, \citenamefont {Clays},\ and\ \citenamefont
  {Persoons}}]{araoka2005int}%
  \BibitemOpen
  \bibfield  {author} {\bibinfo {author} {\bibfnamefont {F.}~\bibnamefont
  {Araoka}}, \bibinfo {author} {\bibfnamefont {T.}~\bibnamefont {Verbiest}},
  \bibinfo {author} {\bibfnamefont {K.}~\bibnamefont {Clays}}, \ and\ \bibinfo
  {author} {\bibfnamefont {A.}~\bibnamefont {Persoons}},\ }\href@noop {}
  {\bibfield  {journal} {\bibinfo  {journal} {Phys.\ Rev.\ A}\ }\textbf
  {\bibinfo {volume} {71}},\ \bibinfo {pages} {055401} (\bibinfo {year}
  {2005})}\BibitemShut {NoStop}%
\bibitem [{\citenamefont {Andrews}(2008)}]{andrews2011str}%
  \BibitemOpen
  \bibfield  {author} {\bibinfo {author} {\bibfnamefont {D.~L.}\ \bibnamefont
  {Andrews}},\ }\href@noop {} {\emph {\bibinfo {title} {Structured light and
  its applications: An introduction to phase-structured beams and nanoscale
  optical forces}}}\ (\bibinfo  {publisher} {Academic Press},\ \bibinfo {year}
  {2008})\BibitemShut {NoStop}%
\bibitem [{\citenamefont {Padgett}\ \emph {et~al.}(2004)\citenamefont
  {Padgett}, \citenamefont {Courtial},\ and\ \citenamefont
  {Allen}}]{padgett2004lig}%
  \BibitemOpen
  \bibfield  {author} {\bibinfo {author} {\bibfnamefont {M.}~\bibnamefont
  {Padgett}}, \bibinfo {author} {\bibfnamefont {J.}~\bibnamefont {Courtial}}, \
  and\ \bibinfo {author} {\bibfnamefont {L.}~\bibnamefont {Allen}},\
  }\href@noop {} {\bibfield  {journal} {\bibinfo  {journal} {Physics Today}\
  }\textbf {\bibinfo {volume} {57}},\ \bibinfo {pages} {35} (\bibinfo {year}
  {2004})}\BibitemShut {NoStop}%
\bibitem [{\citenamefont {Woerdemann}\ \emph {et~al.}(2009)\citenamefont
  {Woerdemann}, \citenamefont {Alpmann},\ and\ \citenamefont
  {Denz}}]{woerdemann2009sel}%
  \BibitemOpen
  \bibfield  {author} {\bibinfo {author} {\bibfnamefont {M.}~\bibnamefont
  {Woerdemann}}, \bibinfo {author} {\bibfnamefont {C.}~\bibnamefont {Alpmann}},
  \ and\ \bibinfo {author} {\bibfnamefont {C.}~\bibnamefont {Denz}},\
  }\href@noop {} {\bibfield  {journal} {\bibinfo  {journal} {Optics Express}\
  }\textbf {\bibinfo {volume} {17}},\ \bibinfo {pages} {22791} (\bibinfo {year}
  {2009})}\BibitemShut {NoStop}%
\bibitem [{\citenamefont {D{\'a}vila~Romero}\ \emph {et~al.}(2002)\citenamefont
  {D{\'a}vila~Romero}, \citenamefont {Andrews},\ and\ \citenamefont
  {Babiker}}]{romero2002qua}%
  \BibitemOpen
  \bibfield  {author} {\bibinfo {author} {\bibfnamefont {L.~C.}\ \bibnamefont
  {D{\'a}vila~Romero}}, \bibinfo {author} {\bibfnamefont {D.~L.}\ \bibnamefont
  {Andrews}}, \ and\ \bibinfo {author} {\bibfnamefont {M.}~\bibnamefont
  {Babiker}},\ }\href@noop {} {\bibfield  {journal} {\bibinfo  {journal} {J.\
  Opt.\ B}\ }\textbf {\bibinfo {volume} {4}},\ \bibinfo {pages} {S66} (\bibinfo
  {year} {2002})}\BibitemShut {NoStop}%
\bibitem [{\citenamefont {Babiker}\ \emph {et~al.}(2002)\citenamefont
  {Babiker}, \citenamefont {Bennett}, \citenamefont {Andrews},\ and\
  \citenamefont {D{\'a}vila~Romero}}]{babiker2002orb}%
  \BibitemOpen
  \bibfield  {author} {\bibinfo {author} {\bibfnamefont {M.}~\bibnamefont
  {Babiker}}, \bibinfo {author} {\bibfnamefont {C.~R.}\ \bibnamefont
  {Bennett}}, \bibinfo {author} {\bibfnamefont {D.~L.}\ \bibnamefont
  {Andrews}}, \ and\ \bibinfo {author} {\bibfnamefont {L.~C.}\ \bibnamefont
  {D{\'a}vila~Romero}},\ }\href@noop {} {\bibfield  {journal} {\bibinfo
  {journal} {Phys.\ Rev.\ Lett.}\ }\textbf {\bibinfo {volume} {89}},\ \bibinfo
  {pages} {143601} (\bibinfo {year} {2002})}\BibitemShut {NoStop}%
\bibitem [{\citenamefont {Quinteiro}\ and\ \citenamefont
  {Tamborenea}(2009{\natexlab{a}})}]{quinteiro2009the}%
  \BibitemOpen
  \bibfield  {author} {\bibinfo {author} {\bibfnamefont {G.~F.}\ \bibnamefont
  {Quinteiro}}\ and\ \bibinfo {author} {\bibfnamefont {P.~I.}\ \bibnamefont
  {Tamborenea}},\ }\href@noop {} {\bibfield  {journal} {\bibinfo  {journal}
  {Europhys.\ Lett.}\ }\textbf {\bibinfo {volume} {85}},\ \bibinfo {pages}
  {47001} (\bibinfo {year} {2009}{\natexlab{a}})}\BibitemShut {NoStop}%
\bibitem [{\citenamefont {Andersen}\ \emph {et~al.}(2006)\citenamefont
  {Andersen}, \citenamefont {Ryu}, \citenamefont {Clad{\'e}}, \citenamefont
  {Natarajan}, \citenamefont {Vaziri}, \citenamefont {Helmerson},\ and\
  \citenamefont {Phillips}}]{andersen2006qua}%
  \BibitemOpen
  \bibfield  {author} {\bibinfo {author} {\bibfnamefont {M.~F.}\ \bibnamefont
  {Andersen}}, \bibinfo {author} {\bibfnamefont {C.}~\bibnamefont {Ryu}},
  \bibinfo {author} {\bibfnamefont {P.}~\bibnamefont {Clad{\'e}}}, \bibinfo
  {author} {\bibfnamefont {V.}~\bibnamefont {Natarajan}}, \bibinfo {author}
  {\bibfnamefont {A.}~\bibnamefont {Vaziri}}, \bibinfo {author} {\bibfnamefont
  {K.}~\bibnamefont {Helmerson}}, \ and\ \bibinfo {author} {\bibfnamefont
  {W.~D.}\ \bibnamefont {Phillips}},\ }\href@noop {} {\bibfield  {journal}
  {\bibinfo  {journal} {Phys.\ Rev.\ Lett.}\ }\textbf {\bibinfo {volume}
  {97}},\ \bibinfo {pages} {170406} (\bibinfo {year} {2006})}\BibitemShut
  {NoStop}%
\bibitem [{\citenamefont {Simula}\ \emph {et~al.}(2008)\citenamefont {Simula},
  \citenamefont {Nygaard}, \citenamefont {Hu}, \citenamefont {Collins},
  \citenamefont {Schneider},\ and\ \citenamefont {M{\o}lmer}}]{simula2008ang}%
  \BibitemOpen
  \bibfield  {author} {\bibinfo {author} {\bibfnamefont {T.~P.}\ \bibnamefont
  {Simula}}, \bibinfo {author} {\bibfnamefont {N.}~\bibnamefont {Nygaard}},
  \bibinfo {author} {\bibfnamefont {S.~X.}\ \bibnamefont {Hu}}, \bibinfo
  {author} {\bibfnamefont {L.~A.}\ \bibnamefont {Collins}}, \bibinfo {author}
  {\bibfnamefont {B.~I.}\ \bibnamefont {Schneider}}, \ and\ \bibinfo {author}
  {\bibfnamefont {K.}~\bibnamefont {M{\o}lmer}},\ }\href@noop {} {\bibfield
  {journal} {\bibinfo  {journal} {Phys.\ Rev.\ A}\ }\textbf {\bibinfo {volume}
  {77}},\ \bibinfo {pages} {015401} (\bibinfo {year} {2008})}\BibitemShut
  {NoStop}%
\bibitem [{\citenamefont {Ueno}\ \emph {et~al.}(2009)\citenamefont {Ueno},
  \citenamefont {Toda}, \citenamefont {Adachi}, \citenamefont {Morita},\ and\
  \citenamefont {Tawara}}]{ueno2009coh}%
  \BibitemOpen
  \bibfield  {author} {\bibinfo {author} {\bibfnamefont {Y.}~\bibnamefont
  {Ueno}}, \bibinfo {author} {\bibfnamefont {Y.}~\bibnamefont {Toda}}, \bibinfo
  {author} {\bibfnamefont {S.}~\bibnamefont {Adachi}}, \bibinfo {author}
  {\bibfnamefont {R.}~\bibnamefont {Morita}}, \ and\ \bibinfo {author}
  {\bibfnamefont {T.}~\bibnamefont {Tawara}},\ }\href@noop {} {\bibfield
  {journal} {\bibinfo  {journal} {Optics Express}\ }\textbf {\bibinfo {volume}
  {17}},\ \bibinfo {pages} {20567} (\bibinfo {year} {2009})}\BibitemShut
  {NoStop}%
\bibitem [{\citenamefont {Shigematsu}\ \emph {et~al.}(2013)\citenamefont
  {Shigematsu}, \citenamefont {Toda}, \citenamefont {Yamane},\ and\
  \citenamefont {Morita}}]{shigematsu2013orb}%
  \BibitemOpen
  \bibfield  {author} {\bibinfo {author} {\bibfnamefont {K.}~\bibnamefont
  {Shigematsu}}, \bibinfo {author} {\bibfnamefont {Y.}~\bibnamefont {Toda}},
  \bibinfo {author} {\bibfnamefont {K.}~\bibnamefont {Yamane}}, \ and\ \bibinfo
  {author} {\bibfnamefont {R.}~\bibnamefont {Morita}},\ }\href@noop {}
  {\bibfield  {journal} {\bibinfo  {journal} {Jpn.\ J.\ Appl.\ Phys.}\ }\textbf
  {\bibinfo {volume} {52}},\ \bibinfo {pages} {08JL08} (\bibinfo {year}
  {2013})}\BibitemShut {NoStop}%
\bibitem [{\citenamefont {W{\"a}tzel}\ \emph {et~al.}(2012)\citenamefont
  {W{\"a}tzel}, \citenamefont {Moskalenko},\ and\ \citenamefont
  {Berakdar}}]{watzel2012pho}%
  \BibitemOpen
  \bibfield  {author} {\bibinfo {author} {\bibfnamefont {J.}~\bibnamefont
  {W{\"a}tzel}}, \bibinfo {author} {\bibfnamefont {A.~S.}\ \bibnamefont
  {Moskalenko}}, \ and\ \bibinfo {author} {\bibfnamefont {J.}~\bibnamefont
  {Berakdar}},\ }\href@noop {} {\bibfield  {journal} {\bibinfo  {journal}
  {Optics Express}\ }\textbf {\bibinfo {volume} {20}},\ \bibinfo {pages}
  {27792} (\bibinfo {year} {2012})}\BibitemShut {NoStop}%
\bibitem [{\citenamefont {Clayburn}\ \emph {et~al.}(2013)\citenamefont
  {Clayburn}, \citenamefont {McCarter}, \citenamefont {Dreiling}, \citenamefont
  {Poelker}, \citenamefont {Ryan},\ and\ \citenamefont
  {Gay}}]{clayburn2013sea}%
  \BibitemOpen
  \bibfield  {author} {\bibinfo {author} {\bibfnamefont {N.~B.}\ \bibnamefont
  {Clayburn}}, \bibinfo {author} {\bibfnamefont {J.~L.}\ \bibnamefont
  {McCarter}}, \bibinfo {author} {\bibfnamefont {J.~M.}\ \bibnamefont
  {Dreiling}}, \bibinfo {author} {\bibfnamefont {M.}~\bibnamefont {Poelker}},
  \bibinfo {author} {\bibfnamefont {D.~M.}\ \bibnamefont {Ryan}}, \ and\
  \bibinfo {author} {\bibfnamefont {T.~J.}\ \bibnamefont {Gay}},\ }\href@noop
  {} {\bibfield  {journal} {\bibinfo  {journal} {Phys.\ Rev.\ {\rm B}}\
  }\textbf {\bibinfo {volume} {87}},\ \bibinfo {pages} {035204} (\bibinfo
  {year} {2013})}\BibitemShut {NoStop}%
\bibitem [{\citenamefont {Quinteiro}\ \emph {et~al.}(2010)\citenamefont
  {Quinteiro}, \citenamefont {Lucero},\ and\ \citenamefont
  {Tamborenea}}]{quinteiro2010ele}%
  \BibitemOpen
  \bibfield  {author} {\bibinfo {author} {\bibfnamefont {G.~F.}\ \bibnamefont
  {Quinteiro}}, \bibinfo {author} {\bibfnamefont {A.~O.}\ \bibnamefont
  {Lucero}}, \ and\ \bibinfo {author} {\bibfnamefont {P.~I.}\ \bibnamefont
  {Tamborenea}},\ }\href@noop {} {\bibfield  {journal} {\bibinfo  {journal}
  {J.\ Phys.\ Cond.\ Matter}\ }\textbf {\bibinfo {volume} {22}},\ \bibinfo
  {pages} {505802} (\bibinfo {year} {2010})}\BibitemShut {NoStop}%
\bibitem [{\citenamefont {Quinteiro}\ and\ \citenamefont
  {Tamborenea}(2010)}]{quinteiro2010twi}%
  \BibitemOpen
  \bibfield  {author} {\bibinfo {author} {\bibfnamefont {G.~F.}\ \bibnamefont
  {Quinteiro}}\ and\ \bibinfo {author} {\bibfnamefont {P.~I.}\ \bibnamefont
  {Tamborenea}},\ }\href@noop {} {\bibfield  {journal} {\bibinfo  {journal}
  {Phys.\ Rev.\ {\rm B}}\ }\textbf {\bibinfo {volume} {82}},\ \bibinfo {pages}
  {125207} (\bibinfo {year} {2010})}\BibitemShut {NoStop}%
\bibitem [{\citenamefont {Quinteiro}(2010)}]{quinteiro2010bel}%
  \BibitemOpen
  \bibfield  {author} {\bibinfo {author} {\bibfnamefont {G.~F.}\ \bibnamefont
  {Quinteiro}},\ }\href@noop {} {\bibfield  {journal} {\bibinfo  {journal}
  {Europhys.\ Lett.}\ }\textbf {\bibinfo {volume} {91}},\ \bibinfo {pages}
  {27002} (\bibinfo {year} {2010})}\BibitemShut {NoStop}%
\bibitem [{\citenamefont {Quinteiro}\ and\ \citenamefont
  {Tamborenea}(2009{\natexlab{b}})}]{quinteiro2009ele1}%
  \BibitemOpen
  \bibfield  {author} {\bibinfo {author} {\bibfnamefont {G.~F.}\ \bibnamefont
  {Quinteiro}}\ and\ \bibinfo {author} {\bibfnamefont {P.~I.}\ \bibnamefont
  {Tamborenea}},\ }\href@noop {} {\bibfield  {journal} {\bibinfo  {journal}
  {Phys.\ Rev.\ {\rm B}}\ }\textbf {\bibinfo {volume} {79}},\ \bibinfo {pages}
  {155450} (\bibinfo {year} {2009}{\natexlab{b}})}\BibitemShut {NoStop}%
\bibitem [{\citenamefont {Quinteiro}\ and\ \citenamefont
  {Berakdar}(2009)}]{quinteiro2009ele2}%
  \BibitemOpen
  \bibfield  {author} {\bibinfo {author} {\bibfnamefont {G.~F.}\ \bibnamefont
  {Quinteiro}}\ and\ \bibinfo {author} {\bibfnamefont {J.}~\bibnamefont
  {Berakdar}},\ }\href@noop {} {\bibfield  {journal} {\bibinfo  {journal}
  {Optics Express}\ }\textbf {\bibinfo {volume} {17}},\ \bibinfo {pages}
  {20465} (\bibinfo {year} {2009})}\BibitemShut {NoStop}%
\bibitem [{\citenamefont {Sbierski}\ \emph {et~al.}(2013)\citenamefont
  {Sbierski}, \citenamefont {Quinteiro},\ and\ \citenamefont
  {Tamborenea}}]{sbierski2013twi}%
  \BibitemOpen
  \bibfield  {author} {\bibinfo {author} {\bibfnamefont {B.}~\bibnamefont
  {Sbierski}}, \bibinfo {author} {\bibfnamefont {G.}~\bibnamefont {Quinteiro}},
  \ and\ \bibinfo {author} {\bibfnamefont {P.}~\bibnamefont {Tamborenea}},\
  }\href@noop {} {\bibfield  {journal} {\bibinfo  {journal} {J.\ Phys.\ Cond.\
  Matter}\ }\textbf {\bibinfo {volume} {25}},\ \bibinfo {pages} {385301}
  (\bibinfo {year} {2013})}\BibitemShut {NoStop}%
\bibitem [{\citenamefont {Padgett}\ and\ \citenamefont
  {Bowman}(2011)}]{padgett2011twe}%
  \BibitemOpen
  \bibfield  {author} {\bibinfo {author} {\bibfnamefont {M.}~\bibnamefont
  {Padgett}}\ and\ \bibinfo {author} {\bibfnamefont {R.}~\bibnamefont
  {Bowman}},\ }\href@noop {} {\bibfield  {journal} {\bibinfo  {journal} {Nature
  Photon.}\ }\textbf {\bibinfo {volume} {5}},\ \bibinfo {pages} {343} (\bibinfo
  {year} {2011})}\BibitemShut {NoStop}%
\bibitem [{\citenamefont {Quinteiro}\ \emph {et~al.}(2013)\citenamefont
  {Quinteiro},\ and\ \citenamefont
  {Kuhn}}]{quinteiro2014lig}%
  \BibitemOpen
  \bibfield  {author} {\bibinfo {author} {\bibfnamefont {G. F.}~\bibnamefont
  {Quinteiro}}, 
  \ and\ \bibinfo {author} {\bibfnamefont {T.}~\bibnamefont {Kuhn}},\
  }\href@noop {} {\bibfield  {journal} {\bibinfo  {journal} {Phys.\ Rev.\ B
  }\ }\textbf {\bibinfo {volume} {90}},\ \bibinfo {pages} {115401}
  (\bibinfo {year} {2014})}\BibitemShut {NoStop}%
\bibitem [{\citenamefont {Woerdemann}\ \emph {et~al.}(2012)\citenamefont
  {Woerdemann}, \citenamefont {Alpmann}, \citenamefont {Esseling},\ and\
  \citenamefont {Denz}}]{woerdemann2012adv}%
  \BibitemOpen
  \bibfield  {author} {\bibinfo {author} {\bibfnamefont {M.}~\bibnamefont
  {Woerdemann}}, \bibinfo {author} {\bibfnamefont {C.}~\bibnamefont {Alpmann}},
  \bibinfo {author} {\bibfnamefont {M.}~\bibnamefont {Esseling}}, \ and\
  \bibinfo {author} {\bibfnamefont {C.}~\bibnamefont {Denz}},\ }\href@noop {}
  {\bibfield  {journal} {\bibinfo  {journal} {Laser Photon. Rev.}\ }\textbf
  {\bibinfo {volume} {7}},\ \bibinfo {pages} {839} (\bibinfo {year}
  {2012})}\BibitemShut {NoStop}%
\bibitem [{\citenamefont {Bozinovic}\ \emph {et~al.}(2013)\citenamefont
  {Bozinovic}, \citenamefont {Yue}, \citenamefont {Ren}, \citenamefont {Tur},
  \citenamefont {Kristensen}, \citenamefont {Huang}, \citenamefont {Willner},\
  and\ \citenamefont {Ramachandran}}]{bozinovic2013ter}%
  \BibitemOpen
  \bibfield  {author} {\bibinfo {author} {\bibfnamefont {N.}~\bibnamefont
  {Bozinovic}}, \bibinfo {author} {\bibfnamefont {Y.}~\bibnamefont {Yue}},
  \bibinfo {author} {\bibfnamefont {Y.}~\bibnamefont {Ren}}, \bibinfo {author}
  {\bibfnamefont {M.}~\bibnamefont {Tur}}, \bibinfo {author} {\bibfnamefont
  {P.}~\bibnamefont {Kristensen}}, \bibinfo {author} {\bibfnamefont
  {H.}~\bibnamefont {Huang}}, \bibinfo {author} {\bibfnamefont {A.~E.}\
  \bibnamefont {Willner}}, \ and\ \bibinfo {author} {\bibfnamefont
  {S.}~\bibnamefont {Ramachandran}},\ }\href@noop {} {\bibfield  {journal}
  {\bibinfo  {journal} {Science}\ }\textbf {\bibinfo {volume} {340}},\ \bibinfo
  {pages} {1545} (\bibinfo {year} {2013})}\BibitemShut {NoStop}%
\bibitem [{\citenamefont {Molina-Terriza}\ \emph {et~al.}(2004)\citenamefont
  {Molina-Terriza}, \citenamefont {Vaziri}, \citenamefont
  {{\v{R}}eh{\'a}{\v{c}}ek}, \citenamefont {Hradil},\ and\ \citenamefont
  {Zeilinger}}]{molina2004tri}%
  \BibitemOpen
  \bibfield  {author} {\bibinfo {author} {\bibfnamefont {G.}~\bibnamefont
  {Molina-Terriza}}, \bibinfo {author} {\bibfnamefont {A.}~\bibnamefont
  {Vaziri}}, \bibinfo {author} {\bibfnamefont {J.}~\bibnamefont
  {{\v{R}}eh{\'a}{\v{c}}ek}}, \bibinfo {author} {\bibfnamefont
  {Z.}~\bibnamefont {Hradil}}, \ and\ \bibinfo {author} {\bibfnamefont
  {A.}~\bibnamefont {Zeilinger}},\ }\href@noop {} {\bibfield  {journal}
  {\bibinfo  {journal} {Phys.\ Rev.\ Lett.}\ }\textbf {\bibinfo {volume}
  {92}},\ \bibinfo {pages} {167903} (\bibinfo {year} {2004})}\BibitemShut
  {NoStop}%
\bibitem [{\citenamefont {Molina-Terriza}\ \emph {et~al.}(2007)\citenamefont
  {Molina-Terriza}, \citenamefont {Torres},\ and\ \citenamefont
  {Torner}}]{molina2007twi}%
  \BibitemOpen
  \bibfield  {author} {\bibinfo {author} {\bibfnamefont {G.}~\bibnamefont
  {Molina-Terriza}}, \bibinfo {author} {\bibfnamefont {J.~P.}\ \bibnamefont
  {Torres}}, \ and\ \bibinfo {author} {\bibfnamefont {L.}~\bibnamefont
  {Torner}},\ }\href@noop {} {\bibfield  {journal} {\bibinfo  {journal} {Nat.
  Phys.}\ }\textbf {\bibinfo {volume} {3}},\ \bibinfo {pages} {305} (\bibinfo
  {year} {2007})}\BibitemShut {NoStop}%
\bibitem [{\citenamefont {D'Ambrosio}\ \emph {et~al.}(2013)\citenamefont
  {D'Ambrosio}, \citenamefont {Spagnolo}, \citenamefont {Del~Re}, \citenamefont
  {Slussarenko}, \citenamefont {Li}, \citenamefont {Kwek}, \citenamefont
  {Marucci}, \citenamefont {Walborn}, \citenamefont {Aolita},\ and\
  \citenamefont {Sciarrino}}]{dambrosio2013pho}%
  \BibitemOpen
  \bibfield  {author} {\bibinfo {author} {\bibfnamefont {V.}~\bibnamefont
  {D'Ambrosio}}, \bibinfo {author} {\bibfnamefont {N.}~\bibnamefont
  {Spagnolo}}, \bibinfo {author} {\bibfnamefont {L.}~\bibnamefont {Del~Re}},
  \bibinfo {author} {\bibfnamefont {S.}~\bibnamefont {Slussarenko}}, \bibinfo
  {author} {\bibfnamefont {Y.}~\bibnamefont {Li}}, \bibinfo {author}
  {\bibfnamefont {L.~C.}\ \bibnamefont {Kwek}}, \bibinfo {author}
  {\bibfnamefont {L.}~\bibnamefont {Marucci}}, \bibinfo {author} {\bibfnamefont
  {S.~P.}\ \bibnamefont {Walborn}}, \bibinfo {author} {\bibfnamefont
  {L.}~\bibnamefont {Aolita}}, \ and\ \bibinfo {author} {\bibfnamefont
  {F.}~\bibnamefont {Sciarrino}},\ }\href@noop {} {\bibfield  {journal}
  {\bibinfo  {journal} {Nat. Commun.}\ }\textbf {\bibinfo {volume} {4}},\
  \bibinfo {pages} {2432} (\bibinfo {year} {2013})}\BibitemShut {NoStop}%
\bibitem [{\citenamefont {Ornigotti}\ and\ \citenamefont
  {Aiello}(2013)}]{ornigotti2013rad}%
  \BibitemOpen
  \bibfield  {author} {\bibinfo {author} {\bibfnamefont {M.}~\bibnamefont
  {Ornigotti}}\ and\ \bibinfo {author} {\bibfnamefont {A.}~\bibnamefont
  {Aiello}},\ }\href@noop {} {\bibfield  {journal} {\bibinfo  {journal} {Optics
  Express}\ }\textbf {\bibinfo {volume} {21}},\ \bibinfo {pages} {15530}
  (\bibinfo {year} {2013})}\BibitemShut {NoStop}%
\bibitem [{\citenamefont {Dorn}\ \emph {et~al.}(2003)\citenamefont {Dorn},
  \citenamefont {Quabis},\ and\ \citenamefont {Leuchs}}]{dorn2003sha}%
  \BibitemOpen
  \bibfield  {author} {\bibinfo {author} {\bibfnamefont {R.}~\bibnamefont
  {Dorn}}, \bibinfo {author} {\bibfnamefont {S.}~\bibnamefont {Quabis}}, \ and\
  \bibinfo {author} {\bibfnamefont {G.}~\bibnamefont {Leuchs}},\ }\href@noop {}
  {\bibfield  {journal} {\bibinfo  {journal} {Phys.\ Rev.\ Lett.}\ }\textbf
  {\bibinfo {volume} {91}},\ \bibinfo {pages} {233901} (\bibinfo {year}
  {2003})}\BibitemShut {NoStop}%
\bibitem [{\citenamefont {Zurita-S{\'a}nchez}\ and\ \citenamefont
  {Novotny}(2002)}]{zurita2002mul}%
  \BibitemOpen
  \bibfield  {author} {\bibinfo {author} {\bibfnamefont {J.~R.}\ \bibnamefont
  {Zurita-S{\'a}nchez}}\ and\ \bibinfo {author} {\bibfnamefont
  {L.}~\bibnamefont {Novotny}},\ }\href@noop {} {\bibfield  {journal} {\bibinfo
   {journal} {J.\ Opt.\ Soc.\ Am.\ B}\ }\textbf {\bibinfo {volume} {19}},\
  \bibinfo {pages} {1355} (\bibinfo {year} {2002})}\BibitemShut {NoStop}%
\bibitem [{\citenamefont {Scully}\ and\ \citenamefont
  {Zubairy}(1997)}]{scully1997qua}%
  \BibitemOpen
  \bibfield  {author} {\bibinfo {author} {\bibfnamefont {M.~O.}\ \bibnamefont
  {Scully}}\ and\ \bibinfo {author} {\bibfnamefont {M.~S.}\ \bibnamefont
  {Zubairy}},\ }\href@noop {} {\emph {\bibinfo {title} {Quantum Optics}}}\
  (\bibinfo  {publisher} {Cambridge University Press, Cambridge},\ \bibinfo
  {year} {1997})\BibitemShut {NoStop}%
\bibitem [{\citenamefont {Cohen-Tannoudji}\ \emph {et~al.}(1989)\citenamefont
  {Cohen-Tannoudji}, \citenamefont {Dupont-Roc},\ and\ \citenamefont
  {Grynberg}}]{cohen1997pho}%
  \BibitemOpen
  \bibfield  {author} {\bibinfo {author} {\bibfnamefont {C.}~\bibnamefont
  {Cohen-Tannoudji}}, \bibinfo {author} {\bibfnamefont {J.}~\bibnamefont
  {Dupont-Roc}}, \ and\ \bibinfo {author} {\bibfnamefont {G.}~\bibnamefont
  {Grynberg}},\ }\href@noop {} {\emph {\bibinfo {title} {Photons and Atoms:
  Introduction to Quantum Electrodynamics}}}\ (\bibinfo  {publisher} {Wiley},\
  \bibinfo {year} {1989})\BibitemShut {NoStop}%
\bibitem [{\citenamefont {Jackson}(2002)}]{jackson2002lorenz}%
  \BibitemOpen
  \bibfield  {author} {\bibinfo {author} {\bibfnamefont {J.~D.}\ \bibnamefont
  {Jackson}},\ }\href@noop {} {\bibfield  {journal} {\bibinfo  {journal} {Am.\
  J.\ Phys.}\ }\textbf {\bibinfo {volume} {70}},\ \bibinfo {pages} {917}
  (\bibinfo {year} {2002})}\BibitemShut {NoStop}%
\bibitem [{\citenamefont {Bassani}\ \emph {et~al.}(1977)\citenamefont
  {Bassani}, \citenamefont {Forney},\ and\ \citenamefont
  {Quattropani}}]{bassani1977cho}%
  \BibitemOpen
  \bibfield  {author} {\bibinfo {author} {\bibfnamefont {F.}~\bibnamefont
  {Bassani}}, \bibinfo {author} {\bibfnamefont {J.~J.}\ \bibnamefont {Forney}},
  \ and\ \bibinfo {author} {\bibfnamefont {A.~A.}\ \bibnamefont
  {Quattropani}},\ }\href@noop {} {\bibfield  {journal} {\bibinfo  {journal}
  {Phys.\ Rev.\ Lett.}\ }\textbf {\bibinfo {volume} {39}},\ \bibinfo {pages}
  {1070} (\bibinfo {year} {1977})}\BibitemShut {NoStop}%
\bibitem [{\citenamefont {Fried}(1973)}]{fried1973vec}%
  \BibitemOpen
  \bibfield  {author} {\bibinfo {author} {\bibfnamefont {Z.}~\bibnamefont
  {Fried}},\ }\href@noop {} {\bibfield  {journal} {\bibinfo  {journal} {Phys.\
  Rev.\ A}\ }\textbf {\bibinfo {volume} {8}},\ \bibinfo {pages} {2835}
  (\bibinfo {year} {1973})}\BibitemShut {NoStop}%
\bibitem [{\citenamefont {Dalgarno}\ and\ \citenamefont
  {Lewis}(1956)}]{dalgarno1956app}%
  \BibitemOpen
  \bibfield  {author} {\bibinfo {author} {\bibfnamefont {A.}~\bibnamefont
  {Dalgarno}}\ and\ \bibinfo {author} {\bibfnamefont {J.~T.}\ \bibnamefont
  {Lewis}},\ }\href@noop {} {\bibfield  {journal} {\bibinfo  {journal} {Proc.
  Phys. Soc. A}\ }\textbf {\bibinfo {volume} {69}},\ \bibinfo {pages} {285}
  (\bibinfo {year} {1956})}\BibitemShut {NoStop}%
\bibitem [{\citenamefont {Dressel}\ and\ \citenamefont
  {Gr{\"u}ner}(2002)}]{dressel2002ele}%
  \BibitemOpen
  \bibfield  {author} {\bibinfo {author} {\bibfnamefont {M.}~\bibnamefont
  {Dressel}}\ and\ \bibinfo {author} {\bibfnamefont {G.}~\bibnamefont
  {Gr{\"u}ner}},\ }\href@noop {} {\emph {\bibinfo {title} {Electrodynamics of
  Solids}}}\ (\bibinfo  {publisher} {Cambridge University Press},\ \bibinfo
  {address} {Cambridge},\ \bibinfo {year} {2002})\BibitemShut {NoStop}%
\bibitem [{\citenamefont {Barnett}\ and\ \citenamefont
  {Loudon}(2010)}]{Barnett2010the}%
  \BibitemOpen
  \bibfield  {author} {\bibinfo {author} {\bibfnamefont {S.~M.}\ \bibnamefont
  {Barnett}}\ and\ \bibinfo {author} {\bibfnamefont {R.}~\bibnamefont
  {Loudon}},\ }\href@noop {} {\bibfield  {journal} {\bibinfo  {journal} {Phil.
  Trans. R. Soc. A}\ }\textbf {\bibinfo {volume} {368}},\ \bibinfo {pages}
  {927} (\bibinfo {year} {2010})}\BibitemShut {NoStop}%
\bibitem [{\citenamefont {Herbst}\ \emph {et~al.}(2003)\citenamefont {Herbst},
  \citenamefont {Glanemann}, \citenamefont {Axt},\ and\ \citenamefont
  {Kuhn}}]{herbst2003ele}%
  \BibitemOpen
  \bibfield  {author} {\bibinfo {author} {\bibfnamefont {M.}~\bibnamefont
  {Herbst}}, \bibinfo {author} {\bibfnamefont {M.}~\bibnamefont {Glanemann}},
  \bibinfo {author} {\bibfnamefont {V.~M.}\ \bibnamefont {Axt}}, \ and\
  \bibinfo {author} {\bibfnamefont {T.}~\bibnamefont {Kuhn}},\ }\href@noop {}
  {\bibfield  {journal} {\bibinfo  {journal} {Phys.\ Rev.\ {\rm B}}\ }\textbf
  {\bibinfo {volume} {67}},\ \bibinfo {pages} {195305} (\bibinfo {year}
  {2003})}\BibitemShut {NoStop}%
\bibitem [{\citenamefont {Rossi}\ and\ \citenamefont
  {Kuhn}(2002)}]{rossi2002the}%
  \BibitemOpen
  \bibfield  {author} {\bibinfo {author} {\bibfnamefont {F.}~\bibnamefont
  {Rossi}}\ and\ \bibinfo {author} {\bibfnamefont {T.}~\bibnamefont {Kuhn}},\
  }\href@noop {} {\bibfield  {journal} {\bibinfo  {journal} {Rev.\ Mod.\
  Phys.}\ }\textbf {\bibinfo {volume} {74}},\ \bibinfo {pages} {895} (\bibinfo
  {year} {2002})}\BibitemShut {NoStop}%
\bibitem [{\citenamefont {Khitrova}\ \emph {et~al.}(1999)\citenamefont
  {Khitrova}, \citenamefont {Gibbs}, \citenamefont {Jahnke}, \citenamefont
  {Kira},\ and\ \citenamefont {Koch}}]{khitrova1999non}%
  \BibitemOpen
  \bibfield  {author} {\bibinfo {author} {\bibfnamefont {G.}~\bibnamefont
  {Khitrova}}, \bibinfo {author} {\bibfnamefont {H.~M.}\ \bibnamefont {Gibbs}},
  \bibinfo {author} {\bibfnamefont {F.}~\bibnamefont {Jahnke}}, \bibinfo
  {author} {\bibfnamefont {M.}~\bibnamefont {Kira}}, \ and\ \bibinfo {author}
  {\bibfnamefont {S.~W.}\ \bibnamefont {Koch}},\ }\href@noop {} {\bibfield
  {journal} {\bibinfo  {journal} {Rev.\ Mod.\ Phys.}\ }\textbf {\bibinfo
  {volume} {71}},\ \bibinfo {pages} {1591} (\bibinfo {year}
  {1999})}\BibitemShut {NoStop}%
\bibitem [{\citenamefont {Reiter}\ \emph {et~al.}(2006)\citenamefont {Reiter},
  \citenamefont {Glanemann}, \citenamefont {Axt},\ and\ \citenamefont
  {Kuhn}}]{reiter2006con}%
  \BibitemOpen
  \bibfield  {author} {\bibinfo {author} {\bibfnamefont {D.}~\bibnamefont
  {Reiter}}, \bibinfo {author} {\bibfnamefont {M.}~\bibnamefont {Glanemann}},
  \bibinfo {author} {\bibfnamefont {V.}~\bibnamefont {Axt}}, \ and\ \bibinfo
  {author} {\bibfnamefont {T.}~\bibnamefont {Kuhn}},\ }\href@noop {} {\bibfield
   {journal} {\bibinfo  {journal} {Phys.\ Rev.\ {\rm B}}\ }\textbf {\bibinfo
  {volume} {73}},\ \bibinfo {pages} {125334} (\bibinfo {year}
  {2006})}\BibitemShut {NoStop}%
\bibitem [{\citenamefont {Reiter}\ \emph {et~al.}(2007)\citenamefont {Reiter},
  \citenamefont {Glanemann}, \citenamefont {Axt},\ and\ \citenamefont
  {Kuhn}}]{reiter2007spatiotemporal}%
  \BibitemOpen
  \bibfield  {author} {\bibinfo {author} {\bibfnamefont {D.}~\bibnamefont
  {Reiter}}, \bibinfo {author} {\bibfnamefont {M.}~\bibnamefont {Glanemann}},
  \bibinfo {author} {\bibfnamefont {V.~M.}\ \bibnamefont {Axt}}, \ and\
  \bibinfo {author} {\bibfnamefont {T.}~\bibnamefont {Kuhn}},\ }\href@noop {}
  {\bibfield  {journal} {\bibinfo  {journal} {Phys.\ Rev.\ {\rm B}}\ }\textbf
  {\bibinfo {volume} {75}},\ \bibinfo {pages} {205327} (\bibinfo {year}
  {2007})}\BibitemShut {NoStop}%
\bibitem [{\citenamefont {Lindberg}\ \emph {et~al.}(1992)\citenamefont
  {Lindberg}, \citenamefont {Binder},\ and\ \citenamefont
  {Koch}}]{lindberg1992theory}%
  \BibitemOpen
  \bibfield  {author} {\bibinfo {author} {\bibfnamefont {M.}~\bibnamefont
  {Lindberg}}, \bibinfo {author} {\bibfnamefont {R.}~\bibnamefont {Binder}}, \
  and\ \bibinfo {author} {\bibfnamefont {S.}~\bibnamefont {Koch}},\ }\href@noop
  {} {\bibfield  {journal} {\bibinfo  {journal} {Phys.\ Rev.\ A}\ }\textbf
  {\bibinfo {volume} {45}},\ \bibinfo {pages} {1865} (\bibinfo {year}
  {1992})}\BibitemShut {NoStop}%
\bibitem [{\citenamefont {Leitenstorfer}\ \emph {et~al.}(1994)\citenamefont
  {Leitenstorfer}, \citenamefont {Lohner}, \citenamefont {Rick}, \citenamefont
  {Leisching}, \citenamefont {Elsaesser}, \citenamefont {Kuhn}, \citenamefont
  {Rossi}, \citenamefont {Stolz},\ and\ \citenamefont
  {Ploog}}]{leitenstorfer1994excitonic}%
  \BibitemOpen
  \bibfield  {author} {\bibinfo {author} {\bibfnamefont {A.}~\bibnamefont
  {Leitenstorfer}}, \bibinfo {author} {\bibfnamefont {A.}~\bibnamefont
  {Lohner}}, \bibinfo {author} {\bibfnamefont {K.}~\bibnamefont {Rick}},
  \bibinfo {author} {\bibfnamefont {P.}~\bibnamefont {Leisching}}, \bibinfo
  {author} {\bibfnamefont {T.}~\bibnamefont {Elsaesser}}, \bibinfo {author}
  {\bibfnamefont {T.}~\bibnamefont {Kuhn}}, \bibinfo {author} {\bibfnamefont
  {F.}~\bibnamefont {Rossi}}, \bibinfo {author} {\bibfnamefont
  {W.}~\bibnamefont {Stolz}}, \ and\ \bibinfo {author} {\bibfnamefont
  {K.}~\bibnamefont {Ploog}},\ }\href@noop {} {\bibfield  {journal} {\bibinfo
  {journal} {Phys.\ Rev.\ {\rm B}}\ }\textbf {\bibinfo {volume} {49}},\
  \bibinfo {pages} {16372} (\bibinfo {year} {1994})}\BibitemShut {NoStop}%
\bibitem [{\citenamefont {B{\'a}nyai}\ \emph {et~al.}(1995)\citenamefont
  {B{\'a}nyai}, \citenamefont {Thoai}, \citenamefont {Reitsamer}, \citenamefont
  {Haug}, \citenamefont {Steinbach}, \citenamefont {Wehner}, \citenamefont
  {Wegener}, \citenamefont {Marschner},\ and\ \citenamefont
  {Stolz}}]{banyai1995exciton}%
  \BibitemOpen
  \bibfield  {author} {\bibinfo {author} {\bibfnamefont {L.}~\bibnamefont
  {B{\'a}nyai}}, \bibinfo {author} {\bibfnamefont {D.~T.}\ \bibnamefont
  {Thoai}}, \bibinfo {author} {\bibfnamefont {E.}~\bibnamefont {Reitsamer}},
  \bibinfo {author} {\bibfnamefont {H.}~\bibnamefont {Haug}}, \bibinfo {author}
  {\bibfnamefont {D.}~\bibnamefont {Steinbach}}, \bibinfo {author}
  {\bibfnamefont {M.}~\bibnamefont {Wehner}}, \bibinfo {author} {\bibfnamefont
  {M.}~\bibnamefont {Wegener}}, \bibinfo {author} {\bibfnamefont
  {T.}~\bibnamefont {Marschner}}, \ and\ \bibinfo {author} {\bibfnamefont
  {W.}~\bibnamefont {Stolz}},\ }\href@noop {} {\bibfield  {journal} {\bibinfo
  {journal} {Phys.\ Rev.\ Lett.}\ }\textbf {\bibinfo {volume} {75}},\ \bibinfo
  {pages} {2188} (\bibinfo {year} {1995})}\BibitemShut {NoStop}%
\bibitem [{\citenamefont {K{\"o}ksal}\ and\ \citenamefont
  {Berakdar}(2012)}]{koksal2012cha}%
  \BibitemOpen
  \bibfield  {author} {\bibinfo {author} {\bibfnamefont {K.}~\bibnamefont
  {K{\"o}ksal}}\ and\ \bibinfo {author} {\bibfnamefont {J.}~\bibnamefont
  {Berakdar}},\ }\href@noop {} {\bibfield  {journal} {\bibinfo  {journal}
  {Phys.\ Rev.\ A}\ }\textbf {\bibinfo {volume} {86}},\ \bibinfo {pages}
  {063812} (\bibinfo {year} {2012})}\BibitemShut {NoStop}%
\bibitem [{\citenamefont {Al-Awfi}\ and\ \citenamefont
  {Babiker}(2000)}]{al2000ato}%
  \BibitemOpen
  \bibfield  {author} {\bibinfo {author} {\bibfnamefont {S.}~\bibnamefont
  {Al-Awfi}}\ and\ \bibinfo {author} {\bibfnamefont {M.}~\bibnamefont
  {Babiker}},\ }\href@noop {} {\bibfield  {journal} {\bibinfo  {journal}
  {Phys.\ Rev.\ A}\ }\textbf {\bibinfo {volume} {61}},\ \bibinfo {pages}
  {033401} (\bibinfo {year} {2000})}\BibitemShut {NoStop}%
\bibitem [{\citenamefont {Klimov}\ \emph {et~al.}(2012)\citenamefont {Klimov},
  \citenamefont {Bloch}, \citenamefont {Ducloy},\ and\ \citenamefont
  {Leite}}]{klimov2012mapping}%
  \BibitemOpen
  \bibfield  {author} {\bibinfo {author} {\bibfnamefont {V.~V.}\ \bibnamefont
  {Klimov}}, \bibinfo {author} {\bibfnamefont {D.}~\bibnamefont {Bloch}},
  \bibinfo {author} {\bibfnamefont {M.}~\bibnamefont {Ducloy}}, \ and\ \bibinfo
  {author} {\bibfnamefont {J.~R.}\ \bibnamefont {Leite}},\ }\href@noop {}
  {\bibfield  {journal} {\bibinfo  {journal} {Phys.\ Rev.\ A}\ }\textbf
  {\bibinfo {volume} {85}},\ \bibinfo {pages} {053834} (\bibinfo {year}
  {2012})}\BibitemShut {NoStop}%
\bibitem [{\citenamefont {Kira}\ \emph {et~al.}(1999)\citenamefont {Kira},
  \citenamefont {Jahnke}, \citenamefont {Hoyer},\ and\ \citenamefont
  {Koch}}]{kira1999qua}%
  \BibitemOpen
  \bibfield  {author} {\bibinfo {author} {\bibfnamefont {M.}~\bibnamefont
  {Kira}}, \bibinfo {author} {\bibfnamefont {F.}~\bibnamefont {Jahnke}},
  \bibinfo {author} {\bibfnamefont {W.}~\bibnamefont {Hoyer}}, \ and\ \bibinfo
  {author} {\bibfnamefont {S.~W.}\ \bibnamefont {Koch}},\ }\href@noop {}
  {\bibfield  {journal} {\bibinfo  {journal} {Progr.\ Quantum Electron.}\
  }\textbf {\bibinfo {volume} {23}},\ \bibinfo {pages} {189} (\bibinfo {year}
  {1999})}\BibitemShut {NoStop}%
\bibitem [{\citenamefont {Kobe}\ and\ \citenamefont
  {Smirl}(1978)}]{kobe1978gau}%
  \BibitemOpen
  \bibfield  {author} {\bibinfo {author} {\bibfnamefont {D.~H.}\ \bibnamefont
  {Kobe}}\ and\ \bibinfo {author} {\bibfnamefont {A.~L.}\ \bibnamefont
  {Smirl}},\ }\href@noop {} {\bibfield  {journal} {\bibinfo  {journal} {Am.\
  J.\ Phys.}\ }\textbf {\bibinfo {volume} {46}},\ \bibinfo {pages} {624}
  (\bibinfo {year} {1978})}\BibitemShut {NoStop}%
\bibitem [{\citenamefont {Saleh}\ and\ \citenamefont
  {Teich}(2007)}]{saleh2007fun}%
  \BibitemOpen
  \bibfield  {author} {\bibinfo {author} {\bibfnamefont {B.~E.~A.}\
  \bibnamefont {Saleh}}\ and\ \bibinfo {author} {\bibfnamefont {M.~C.}\
  \bibnamefont {Teich}},\ }\href@noop {} {\emph {\bibinfo {title} {Fundamentals
  of of Photonics}}}\ (\bibinfo  {publisher} {Wiley},\ \bibinfo {year}
  {2007})\BibitemShut {NoStop}%
\bibitem [{\citenamefont {J{\'a}uregui}(2004)}]{jauregui2004rot}%
  \BibitemOpen
  \bibfield  {author} {\bibinfo {author} {\bibfnamefont {R.}~\bibnamefont
  {J{\'a}uregui}},\ }\href@noop {} {\bibfield  {journal} {\bibinfo  {journal}
  {Phys.\ Rev.\ A}\ }\textbf {\bibinfo {volume} {70}},\ \bibinfo {pages}
  {033415} (\bibinfo {year} {2004})}\BibitemShut {NoStop}%
\bibitem{VanEnk94} S. J. van Enk and G. Nienhuis 1994 Europhys. Lett. {\bf 25} 497.
\bibitem [{\citenamefont {Volke-Sepulveda}\ \emph {et~al.}(2002)\citenamefont
  {Volke-Sepulveda}, \citenamefont {Garc{\'e}s-Ch{\'a}vez}, \citenamefont
  {Ch{\'a}vez-Cerda}, \citenamefont {Arlt},\ and\ \citenamefont
  {Dholakia}}]{volke2002orb}%
  \BibitemOpen
  \bibfield  {author} {\bibinfo {author} {\bibfnamefont {K.}~\bibnamefont
  {Volke-Sepulveda}}, \bibinfo {author} {\bibfnamefont {V.}~\bibnamefont
  {Garc{\'e}s-Ch{\'a}vez}}, \bibinfo {author} {\bibfnamefont {S.}~\bibnamefont
  {Ch{\'a}vez-Cerda}}, \bibinfo {author} {\bibfnamefont {J.}~\bibnamefont
  {Arlt}}, \ and\ \bibinfo {author} {\bibfnamefont {K.}~\bibnamefont
  {Dholakia}},\ }\href@noop {} {\bibfield  {journal} {\bibinfo  {journal} {J.\
  Opt.\ B}\ }\textbf {\bibinfo {volume} {4}},\ \bibinfo {pages} {S82} (\bibinfo
  {year} {2002})}\BibitemShut {NoStop}%
\bibitem [{\citenamefont {Friese}\ \emph {et~al.}(1998)\citenamefont {Friese},
  \citenamefont {Nieminen}, \citenamefont {Heckenberg},\ and\ \citenamefont
  {Rubinsztein-Dunlop}}]{friese1998opt}%
  \BibitemOpen
  \bibfield  {author} {\bibinfo {author} {\bibfnamefont {M.~E.~J.}\
  \bibnamefont {Friese}}, \bibinfo {author} {\bibfnamefont {T.~A.}\
  \bibnamefont {Nieminen}}, \bibinfo {author} {\bibfnamefont {N.~R.}\
  \bibnamefont {Heckenberg}}, \ and\ \bibinfo {author} {\bibfnamefont
  {H.}~\bibnamefont {Rubinsztein-Dunlop}},\ }\href@noop {} {\bibfield
  {journal} {\bibinfo  {journal} {Nature}\ }\textbf {\bibinfo {volume} {394}},\
  \bibinfo {pages} {348} (\bibinfo {year} {1998})}\BibitemShut {NoStop}%
\bibitem [{\citenamefont {Nienhuis}\ and\ \citenamefont
  {Visser}(2004)}]{nienhuis2004ang}%
  \BibitemOpen
  \bibfield  {author} {\bibinfo {author} {\bibfnamefont {G.}~\bibnamefont
  {Nienhuis}}\ and\ \bibinfo {author} {\bibfnamefont {J.}~\bibnamefont
  {Visser}},\ }\href@noop {} {\bibfield  {journal} {\bibinfo  {journal} {J.\
  Opt.\ A}\ }\textbf {\bibinfo {volume} {6}},\ \bibinfo {pages} {S248}
  (\bibinfo {year} {2004})}\BibitemShut {NoStop}%
\bibitem [{\citenamefont {Tabosa}\ and\ \citenamefont
  {Petrov}(1999)}]{tabosa1999opt}%
  \BibitemOpen
  \bibfield  {author} {\bibinfo {author} {\bibfnamefont {J.~W.~R.}\
  \bibnamefont {Tabosa}}\ and\ \bibinfo {author} {\bibfnamefont {D.~V.}\
  \bibnamefont {Petrov}},\ }\href@noop {} {\bibfield  {journal} {\bibinfo
  {journal} {Phys.\ Rev.\ Lett.}\ }\textbf {\bibinfo {volume} {83}},\ \bibinfo
  {pages} {4967} (\bibinfo {year} {1999})}\BibitemShut {NoStop}%
\bibitem [{\citenamefont {Iketaki}\ \emph {et~al.}(2007)\citenamefont
  {Iketaki}, \citenamefont {Watanabe}, \citenamefont {Bokor},\ and\
  \citenamefont {Fujii}}]{iketaki2007inv}%
  \BibitemOpen
  \bibfield  {author} {\bibinfo {author} {\bibfnamefont {Y.}~\bibnamefont
  {Iketaki}}, \bibinfo {author} {\bibfnamefont {T.}~\bibnamefont {Watanabe}},
  \bibinfo {author} {\bibfnamefont {N.}~\bibnamefont {Bokor}}, \ and\ \bibinfo
  {author} {\bibfnamefont {M.}~\bibnamefont {Fujii}},\ }\href@noop {}
  {\bibfield  {journal} {\bibinfo  {journal} {Opt.\ Lett.}\ }\textbf {\bibinfo
  {volume} {32}},\ \bibinfo {pages} {2357} (\bibinfo {year}
  {2007})}\BibitemShut {NoStop}%
\bibitem [{\citenamefont {Bokor}\ \emph {et~al.}(2005)\citenamefont {Bokor},
  \citenamefont {Iketaki}, \citenamefont {Watanabe},\ and\ \citenamefont
  {Fujii}}]{bokor2005inv}%
  \BibitemOpen
  \bibfield  {author} {\bibinfo {author} {\bibfnamefont {N.}~\bibnamefont
  {Bokor}}, \bibinfo {author} {\bibfnamefont {Y.}~\bibnamefont {Iketaki}},
  \bibinfo {author} {\bibfnamefont {T.}~\bibnamefont {Watanabe}}, \ and\
  \bibinfo {author} {\bibfnamefont {M.}~\bibnamefont {Fujii}},\ }\href@noop {}
  {\bibfield  {journal} {\bibinfo  {journal} {Optics Express}\ }\textbf
  {\bibinfo {volume} {13}},\ \bibinfo {pages} {10440} (\bibinfo {year}
  {2005})}\BibitemShut {NoStop}%
\bibitem [{\citenamefont {Heeres}\ \emph {et~al.}(2014)\citenamefont {Heeres},\
  and\ \citenamefont {Zwiller}}]{heeres2014sub}%
  \BibitemOpen
  \bibfield  {author} {\bibinfo {author} {\bibfnamefont {R.~W.}~\bibnamefont
  {Heeres}},\ and\
  \bibinfo {author} {\bibfnamefont {V.}~\bibnamefont {Zwiller}},\ }\href@noop {}
  {\bibfield  {journal} {\bibinfo  {journal} {Nano letters}\ }\textbf
  {\bibinfo {volume} {14}},\ \bibinfo {pages} {4598} (\bibinfo {year}
  {2014})}\BibitemShut {NoStop}%
\bibitem [{\citenamefont {Wolf}(1959)}]{wolf1959ele}%
  \BibitemOpen
  \bibfield  {author} {\bibinfo {author} {\bibfnamefont {E.}~\bibnamefont
  {Wolf}},\ }\href@noop {} {\bibfield  {journal} {\bibinfo  {journal} {Proc.\
  Roy.\ Soc.\ London, Ser. A}\ }\textbf {\bibinfo {volume} {253}},\ \bibinfo
  {pages} {349} (\bibinfo {year} {1959})}\BibitemShut {NoStop}%
\bibitem [{\citenamefont {Monteiro}\ \emph {et~al.}(2009)\citenamefont
  {Monteiro}, \citenamefont {Neto},\ and\ \citenamefont
  {Nussenzveig}}]{monteiro2009ang}%
  \BibitemOpen
  \bibfield  {author} {\bibinfo {author} {\bibfnamefont {P.~B.}\ \bibnamefont
  {Monteiro}}, \bibinfo {author} {\bibfnamefont {P.~A.~M.}\ \bibnamefont
  {Neto}}, \ and\ \bibinfo {author} {\bibfnamefont {H.~M.}\ \bibnamefont
  {Nussenzveig}},\ }\href@noop {} {\bibfield  {journal} {\bibinfo  {journal}
  {Phys.\ Rev.\ A}\ }\textbf {\bibinfo {volume} {79}},\ \bibinfo {pages}
  {033830} (\bibinfo {year} {2009})}\BibitemShut {NoStop}%
\bibitem [{\citenamefont {Goldstein}(1980)}]{goldstein1962cla}%
  \BibitemOpen
  \bibfield  {author} {\bibinfo {author} {\bibfnamefont {H.}~\bibnamefont
  {Goldstein}},\ }\href@noop {} {\emph {\bibinfo {title} {Classical
  mechanics}}}\ (\bibinfo  {publisher} {Addison-Wesley, second edition},\
  \bibinfo {year} {1980})\BibitemShut {NoStop}%
\end{thebibliography}

%


\end{document}